\documentclass[reprint, aps, prl, groupedaddress, twocolumn, showkeys, nobibnotes, noeprint, nofootinbib]{revtex4-2}

\usepackage{graphicx}
\usepackage{dcolumn}
\usepackage{amsmath}
\usepackage{bm}
\usepackage[utf8]{inputenc}
\usepackage{mathtools}
\usepackage{pgffor}
\usepackage{xcolor}
\usepackage{BOONDOX-cal}
\usepackage{braket}
\usepackage{etaremune}
\usepackage{xcolor}
\usepackage[colorlinks]{hyperref}
\usepackage[mediumspace,mediumqspace,squaren]{SIunits}
\usepackage{upgreek}
\usepackage{mathtools}
\usepackage{verbatim}
\usepackage{enumitem}
\usepackage{setspace}
\definecolor{mygreen}{RGB}{0, 88, 39}
\definecolor{myblue}{RGB}{46, 48, 146}

\hypersetup{
colorlinks=true,
linkcolor=myblue,
filecolor=magenta, 
urlcolor=myblue,
citecolor=myblue
}


\begin{document}

\title{{Fractional quantum Hall state at $\nu = 1/2$ with energy gap up to 6 K, and possible transition from one- to two-component state}}
\date{\today}
\author{Siddharth Kumar Singh, Chengyu Wang, Adbhut Gupta, Kirk W. Baldwin, Loren N. Pfeiffer, and Mansour Shayegan} 

\affiliation{Department of Electrical and Computer Engineering, Princeton University, Princeton, New Jersey 08544, USA}

\let\oldmaketitle\maketitle
\renewcommand\maketitle{{\bfseries\boldmath\oldmaketitle}}


\begin{abstract}{The fractional quantum Hall state (FQHS) observed in the lowest Landau level at filling factor $\nu=1/2$ in wide quantum wells has been enigmatic for decades because the two-dimensional electron system (2DES) has a bilayer charge distribution but with significant interlayer tunneling. {Of particular interest is whether the 1/2 FQHS in this system has a one-component (1C) or two-component (2C) origin; these are typically identified as the Pfaffian (non-Abelian) or the $\Psi_{331}$ (Abelian) FQHSs, respectively. We report here our experimental study of the evolution of the correlated states of an ultrahigh-quality 2DES confined to a 72.5-nm-wide GaAs quantum well. At the lowest densities, the 2DES displays only odd-denominator FQHSs, and the ground state at $\nu = 1/2$ is a composite fermion Fermi sea. As the density is increased, a FQHS emerges at $\nu = 1/2$, and becomes very strong. In a finite density range where the 1/2 FQHS is strongest, we also observe its daughter FQHSs at $\nu = 8/17$ and 7/13, consistent with the theoretically expected daughter states of a Pfaffian 1/2 FQHS. At the highest densities, the 2DES becomes 2C, signaled by the emergence of a bilayer Wigner crystal state and the transitions of FQHSs flanking $\nu=1/2$. The 1/2 FQHS remains robust near this transition and, notably, its charge transport energy gap exhibits an \textit{upward} cusp with a maximum value of about 6 K on the 1C side of the transition; this is the largest gap reported for any even-denominator FQHS. Our observation of the transition of the 2DES ground states near $\nu=1/2$ to 2C states at high densities, and our measurements of the robustness of the 1/2 FQHS against charge distribution asymmetry, suggest that the 1/2 FQHS also makes a transition from 1C to 2C. Such a transition from a non-Abelian to Abelian state can open avenues for topological quantum information and quantum criticality.}}

\end{abstract}

\maketitle

In low-disorder, interacting, two-dimensional electron systems (2DESs), the composite fermions (CFs) at a half-filled Landau level (LL) can undergo pairing to exhibit even-denominator fractional quantum Hall states (FQHSs). This was first observed at LL filling factor $\nu = 5/2$ in the $N = 1$ LL \cite{Willett.PRL.1987}. The 5/2 FQHS is widely believed to host non-Abelian anyons \cite{Nayak.RevModPhys.2008, Banerjee.Nature.2018, Willett.PRX.2023} and is most likely characterized by one of the three wavefunctions, Pfaffian \cite{MooreRead.NuclPhysB.1990}, anti-Pfaffian \cite{Lee.PRL.2007, Levin.PRL.2007} or PH-Pfaffian \cite{Son.PRX.2015}. The CF pairing is facilitated by the weaker electronic repulsion in the $N = 1$ LL because of the node wavefunction \cite{Nayak.RevModPhys.2008, Scarola.Nature.2000}. Non-Abelian anyons have potential use for highly fault-tolerant quantum technologies \cite{Sarma.PRL.2005, Nayak.RevModPhys.2008}, leading to significant interest in even-denominator FQHSs. 

In the $N = 0$ LL, even-denominator FQHSs were also observed at $\nu = 1/2$ in 2DESs confined to wide GaAs quantum wells (QWs) \cite{Suen.PRL.1992, Suen2.PRL.1992, Suen.PRL.1994} and double QWs \cite{Eisenstein.PRL.1992}. The 2DES in a wide QW has a bilayer charge distribution but with significant interlayer tunneling \cite{Suen.PRB.1991}. This has led to several experimental \cite{Suen.PRL.1992, Suen2.PRL.1992, Suen.PRL.1994, Shabani.PRL.2009, Shabani2.PRL.2009, Shabani.PRB.2013, Mueed.PRL.2015, Mueed.PRL.2016, Hasdemir.PRB.2017, Singh.NatPhys.2024} and theoretical \cite{Greiter.PRB.1992, Greiter.NucPhysB.1992, He.PRB.1993, Nomura.JPSJ.2004, Peterson.PRB.2010, Thiebaut.PRB.2015, Zhu.PRB.2016, Sharma.PRB.2024} works debating whether the 1/2 FQHS in wide GaAs QWs is a one-component (1C), non-Abelian state stabilized by large tunneling, or a two-component (2C), Abelian $(\Psi_{331})$ state \cite{Halperin.Helv.Phys.Acta.1983} which is stable in a bilayer system with negligible tunneling. Some of the earlier studies \cite{He.PRB.1993, Suen.PRL.1994, Peterson.PRB.2010, Thiebaut.PRB.2015} argued for a 2C origin. However, direct measurements of the CF Fermi sea wavevector \cite{Mueed.PRL.2015, Mueed.PRL.2016}, and some theoretical calculations \cite{Greiter.PRB.1992, Greiter.NucPhysB.1992, Nomura.JPSJ.2004, Zhu.PRB.2016, Sharma.PRB.2024} favor a 1C, non-Abelian origin. Most recently, the observation of the daughter states of the 1/2 FQHS at $\nu = 8/17$ and 7/13 was reported in 2DESs with appropriate parameters, confined to ultrahigh-quality wide GaAs QWs \cite{Singh.NatPhys.2024}. Insofar as these are precisely the theoretically expected fillings for the daughter states of the Pfaffian 1/2 FQHS \cite{Levin.PRB.2009, Yutushui.PRB.2024, Zheltonozhskii.PRB.2024}, the observations strongly suggest a 1C, non-Abelian origin. This is also corroborated by very recent theoretical work \cite{Sharma.PRB.2024} which calculates a phase boundary between the CF Fermi sea and a 1/2 FQHS using a variational wavefunction that is in the same universality class as the Pfaffian state.

\begin{figure*}[t!]
\hypertarget{fig1}{}
\includegraphics[width=2\columnwidth]{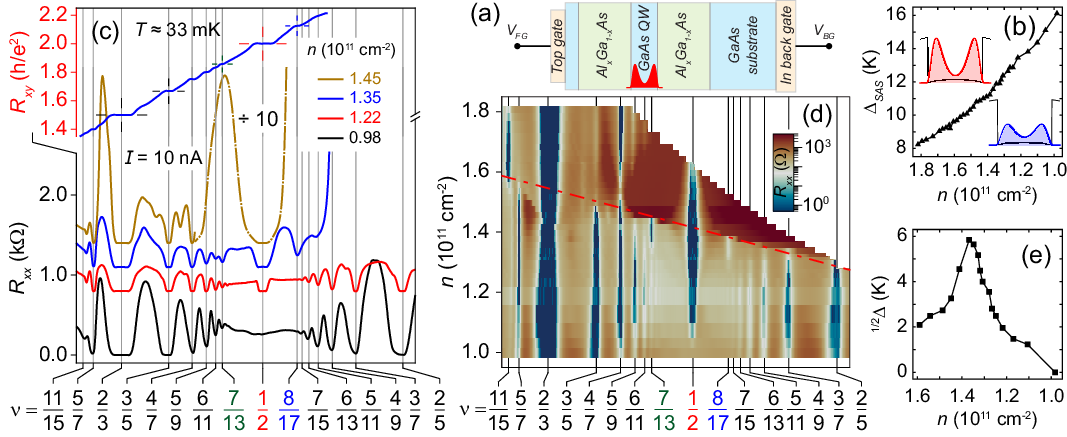}
\caption{\label{fig:fig1}(a) Sample schematic. (b) $\Delta_{SAS}$ extracted from the Shubnikov-de Haas oscillations at low \textit{B}-field. Self-consistent Hartree calculations of the charge distribution and potential at \textit{n} = 1.82 and 1.00 in a 72.5-nm-wide GaAs QW are shown as top and bottom insets, respectively. (c) Characteristic longitudinal resistance $R_{xx}$ traces which coarsely capture the most notable transitions such as the CF Fermi sea to FQHS at $\nu = 1/2$ (red trace), the emergence of the daughter states of the 1/2 FQHS at $\nu = 8/17$ and 7/13 (blue trace), and the bilayer Wigner crystal flanking $\nu = 1/2$ at the highest densities (brown trace). Also shown is the Hall resistance $(R_{xy})$ at \textit{n} = 1.35, displaying well quantized plateaus for the $\nu = 1/2$, 8/17 and 7/13 FQHSs. (d) Color-scale phase diagram of the 2DES in the \textit{N} = 0 LL, rich with correlated electronic solid and liquid phases. {Transitions of the odd-denominator FQHSs and the emergence of a bilayer Wigner crystal signal the 1C to 2C transition of the 2DES as \textit{n} increases (indicated by the red dashed-dotted line, a guide to the eye).} (e) Energy gap of the 1/2 FQHS ($^{1/2}\Delta$) as a function of \textit{n}, peaking to about 6 K.}
\end{figure*}

{We report here our latest findings on the evolution of the correlated states in a wide GaAs QW as a function of interlayer tunneling which we tune by changing the electron density (\textit{n}) in the QW \cite{Suen.PRB.1991, Suen2.PRL.1992, Suen.PRL.1994, Manoharan.PRL.1996, Manoharan.PRL.1997, Shabani.PRB.2013, Singh.NatPhys.2024}.} Figure \hyperlink{fig1}{1(a)} shows a simple schematic of our sample with front and back gates to control both \textit{n} and charge distribution symmetry. Shown in Fig. \hyperlink{fig1}{1(b)} are the interlayer tunneling values determined from the Fourier transforms of the low-field Shubnikov-de Haas oscillations \cite{Suen.PRL.1992, Suen2.PRL.1992, Suen.PRL.1994, Manoharan.PRL.1996, Shayegan.Review.LesHouches.1999}; see Supplemental Material (SM) Section I \cite{SM.2025}. The parameter $\Delta_{SAS}$, which is the energy difference between the QW's symmetric and antisymmetric electric subbands, directly measures the interlayer tunneling of a symmetric charge distribution. As \textit{n} is raised, the electrons repel each other, forcing them towards the QW's walls and resulting in a bilayer charge distribution. This also raises the potential near the QW's center thus reducing the interlayer tunneling \cite{Suen.PRL.1992, Suen2.PRL.1992, Suen.PRL.1994, Manoharan.PRL.1996, Shayegan.Review.LesHouches.1999}; see insets in Fig. \hyperlink{fig1}{1(b)}, also SM Section IV \cite{SM.2025}.

Our sample is a modulation-doped, 72.5-nm-wide, GaAs QW with as-grown \textit{n} = 1.16 (in units of $10^{11}$ cm$^{-2}$ used throughout this manuscript) and mobility $\mu \simeq 10\times10^{6}$ cm\textsuperscript{2}/Vs at \textit{T} = 0.3 K, an order of magnitude larger than the samples in which the 1/2 FQHS was first observed \cite{Suen.PRL.1992}. This improvement is evident from the $R_{xx}$ traces shown in Fig. \hyperlink{fig1}{1(c)}. The traces are plotted as a function of 1/$\nu$ with the x-axis tick marks displaying the expected positions of the various FQHSs. The lowest two traces at \textit{n} = 0.98 and 1.22 show the transition of the CF Fermi sea to a robust FQHS at $\nu = 1/2$. The 1C nature of the 1/2 FQHS can be inferred from the numerous high-order, odd-denominator Jain states \cite{Jain.composite.fermions.2007} flanking $\nu = 1/2$. At \textit{n} = 1.35, the daughter states of the 1/2 FQHS emerge at $\nu = 8/17$ and 7/13 as anomalously deep minima in $R_{xx}$ accompanied by well-developed plateaus in the $R_{xy}$ \cite{Singh.NatPhys.2024}; see top trace in Fig. \hyperlink{fig1}{1(c)}. The values of $\nu$ for these daughter states are consistent with those theoretically expected for the Pfaffian state \cite{Levin.PRB.2009, Yutushui.PRB.2024, Zheltonozhskii.PRB.2024}. Another major feature seen in Figs. \hyperlink{fig1}{1(c,d)} is an insulating phase which first emerges at small fillings ($\nu<8/17$ at \textit{n} = 1.35) and eventually surrounds the 1/2 FQHS at \textit{n} = 1.45. Previous transport \cite{Suen2.PRL.1992, Manoharan.PRL.1996} and microwave resonance studies \cite{Hatke.PRB.2017} identify this insulating phase as a pinned \textit{bilayer} Wigner crystal. 

The 2DES in wide GaAs QWs is rich with several high-order FQHSs at various $\nu$ as \textit{n} is varied between $0.98\lesssim n\lesssim 1.82$, summarized in Fig. \hyperlink{fig1}{1(d)}; see SM Section II \cite{SM.2025} for a complete set of $R_{xx}$ traces \cite{SM.2025}. {Evidence for a 1C to 2C transition of the 2DES can be seen in Fig. \hyperlink{fig1}{1(d)} at states which are now accessible thanks to the much improved quality of our sample; these will be discussed at length later in the manuscript.}

Figure \hyperlink{fig1}{1(e)} manifests the robustness of the 1/2 FQHS in the form of its energy gap ($^{1/2}\Delta$) plotted as a function of \textit{n}. $^{1/2}\Delta$ is extracted by measuring $R_{xx}$ at $\nu=1/2$ for $0.03 \lesssim T \lesssim 0.4$ K and fitting it to the expression $R_{xx} \propto exp(-^{1/2}\Delta/2k_BT)$; see SM Section III \cite{SM.2025}. The most striking feature here is that the 1/2 FQHS at its maximum strength has an energy gap of about 6 K ($n = 1.37$). On either direction of the peak energy gap, $^{1/2}\Delta$ falls off sharply, by $\simeq50 \%$ to about 3 K at \textit{n} = 1.28 and 1.45, which correspond to less than $7 \%$ change in \textit{n}.

\begin{figure*}[t!]
\hypertarget{fig3}{}
\includegraphics[width=2\columnwidth]{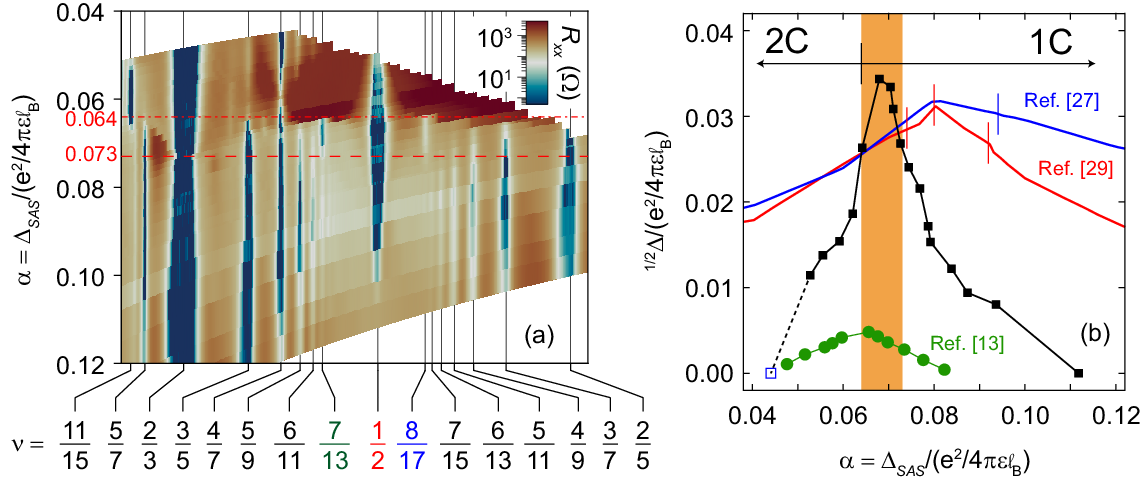}
\caption{\label{fig:fig3}Data of Figs. \protect\hyperlink{fig1}{1(d, e)} are replotted in (a) and (b) respectively using the parameter $\alpha = \Delta_{SAS}/(e^2/4\pi \epsilon \ell_B)$, the interlayer tunneling normalized to the Coulomb energy. (a) The daughter states of a 1C 1/2 FQHS emerge at $0.064 \lesssim \alpha \lesssim 0.073$ while the 1C to 2C transition of the 2DES near $\nu = 1/2$ occurs at $\alpha \lesssim 0.064$. (b) The measured $^{1/2}\Delta$ (black squares), normalized to the Coulomb energy {(see note EM4)}, are about 8 times larger than those reported in a sample with much lower mobility (green circles) \cite{Suen.PRL.1994}. The orange band near $\alpha \simeq 0.07$ denotes the region of stability of the daughter 8/17 and 7/13 FQHSs, and straddles the peak $^{1/2}\Delta$ quite symmetrically. Also plotted are theoretical calculations for $^{1/2}\Delta$ in Ref. \cite{Peterson.PRB.2010} (blue) and Ref. \cite{Zhu.PRB.2016} (red). Note that both sets of calculations predict a boundary near the maximum $^{1/2}\Delta$ between the Pfaffian and $\Psi_{331}$ phases as $\alpha$ is reduced. The vertical lines denote the experimental (black, at $\alpha \simeq 0.064$) 1C-2C transition of the 2DES, and theoretical (red and blue) predictions of the 1C-2C transition of the 1/2 FQHS; see SM Section VI \cite{SM.2025} for details.} 
\end{figure*}

It is illuminating to examine the evolution of the correlated states in the lowest LL and their various transitions as a function of the parameter $\alpha = {\Delta_{SAS}}/({e^2/4 \pi \epsilon \ell_B})$ instead of \textit{n}, as shown in Fig. \hyperlink{fig3}{2}; $\epsilon$ is the GaAs dielectric constant, and $\ell_B = \sqrt{\hbar/eB}$ is the magnetic length. The parameter $\alpha$ is the interlayer tunneling normalized to the Coulomb energy and is used to parameterize theoretical calculations and their comparison with experimental data \cite{Peterson.PRB.2010, Zhu.PRB.2016}. Similarly, $^{1/2}\Delta$ is expressed in units of the Coulomb energy. {Figure \hyperlink{fig3}{2(a)} makes it clear that the 2DES makes a 1C to 2C transition near $\nu=1/2$ at $\alpha \simeq 0.064$.} This is heralded by the emergence of the bilayer Wigner crystal insulating phases \cite{Suen2.PRL.1992, Manoharan.PRL.1996, Hatke.PRB.2017} on the flanks of $\nu = 1/2$ for $\alpha < 0.064$, and the disappearance of the \textit{odd-numerator} FQHSs at 3/7, 5/11, 5/9, 3/5, and 5/7 when $\alpha < 0.064$ {[see note EM1 in End Matter (EM)]}. In contrast, the \textit{even-numerator} FQHS at $\nu = 4/7$ {is weakest at $\alpha \simeq 0.06$, also signaling a 1C to 2C transition {(see notes EM2 and EM3)}.}

{It is evident in Fig. \hyperlink{fig3}{2} that, while the 2DES is undergoing a 1C to 2C transition, the 1/2 FQHS remains strong and in fact has its maximum energy gap near this transition [Fig. \hyperlink{fig3}{2(b)}].} Also significant is that the FQHSs at $\nu = 8/17$ and 7/13, which are the predicted, simplest daughter states of a Pfaffian 1/2 FQHS \cite{Levin.PRB.2009, Yutushui.PRB.2024, Zheltonozhskii.PRB.2024}, are seen in the range $0.064 < \alpha < 0.073$ where the 2DES exhibits 1C FQHSs while also displaying the largest energy gap [see orange shaded region in Fig. \hyperlink{fig3}{2(b)}]. These observations provide strong evidence that the 1/2 FQHS we observe in the range $\alpha > 0.064$ has a 1C origin. {Moreover, Fig. \hyperlink{fig3}{2(a)} clearly shows that the 1/2 FQHS is observed even when $\alpha < 0.064$ and the 2DES has entered the 2C regime. The evolution of the interaction landscape of the 2DES raises the interesting question of a possible 1C to 2C transition of the 1/2 FQHS itself.}

{The susceptibility of the 1/2 FQHS to charge distribution asymmetry is another microscopic property which can shed light on the nature of the 1/2 FQHS.} We probe this susceptibility by appropriately biasing the back and front gates to transfer charge from one side of the QW to the other side while keeping \textit{n} in the QW constant; we quantify this charge transfer by $\Delta n/n$ where $\Delta n$ is the difference of density between two sides of the QW determined from the measured variation of \textit{n} with gate biases. In Fig. \hyperlink{fig2}{3(a)} we show examples of self-consistently calculated charge distributions for $n=1.45$ and 1.28 for the symmetric (upper panels) and asymmetric (lower panels) cases, as indicated. Our measurements show that, regardless of $n$, the 1/2 FQHS monotonically becomes weaker and eventually turns into a compressible CF Fermi sea beyond a critical charge distribution asymmetry $(\Delta n/n)_C$ that depends on $n$. Data for $n = 1.17$ are shown in Fig. \hyperlink{fig2}{3(b)}, revealing the destabilization of the 1/2 FQHS at $(\Delta n/n)_C \simeq 18 \%$.

The measured $(\Delta n/n)_C$ at different densities are plotted in Fig. \hyperlink{fig2}{3(c)} (red cirles, right y-axis), and contrasted against $^{1/2}\Delta$ (black squares, left y-axis). {We find that the trend of $(\Delta n/n)_C$ is correlated to that of $^{1/2}\Delta$ as a function of \textit{n}. However, smaller values of $(\Delta n/n)_C$ are required to destabilize the 1/2 FQHS on the high-density side of the peak $^{1/2}\Delta$, as compared to the low-density side.} For example, when the charge distribution is symmetric, the 1/2 FQHS has the same strength ($^{1/2}\Delta \simeq 3.3$ K) at $n = 1.45$ and 1.28. However, $(\Delta n/n)_C \simeq 13 \%$ for $n=1.45$ is significantly smaller than $(\Delta n/n)_C \simeq 27 \%$ for $n=1.28$. {Considering the intuitive expectation that a 2C 1/2 FQHS, e.g. the $\Psi_{331}$ state, would be less robust against charge distribution asymmetry compared to a 1C state such as the Pfaffian \cite{Peterson2.PRB.2010} or a Pfaffian-like state \cite{Sharma.PRB.2024}, our observation suggests that the 1/2 FQHS we see at high densities is a 2C state. }

A comparison of our data to those reported in Ref. \cite{Suen.PRL.1994}, which studied the evolution of the 1/2 FQHS in a 77-nm-wide-GaAs QW, is instructive. In Fig. \hyperlink{fig3}{2(b)} we have included the $^{1/2}\Delta$ vs $\alpha$ data (open circles) reported in Ref. \cite{Suen.PRL.1994}. The data are qualitatively consistent with our new results in that $^{1/2}\Delta$ vs $\alpha$ exhibits a peak, and the peak position (in $\alpha$) is in fact very close to what we observe. However, there are several important differences; we attribute these to the much higher quality of our sample which has about 10 times larger mobility (SM Section VII \cite{SM.2025}). First, the measured $^{1/2}\Delta$ are about 8 times larger in the new sample, reflecting the much less significant role of disorder. Second, the new sample exhibits many more FQHSs on the flanks of the $\nu = 1/2$, including daughter states of the 1/2 FQHS at $\nu = 8/17$ and 7/13 when the 1/2 FQHS is strong. These collectively allow us to observe a clear 1C to 2C transition of the 2DES (below $\alpha \simeq 0.064$). Third, in Ref. \cite{Suen.PRL.1994}, typical $(\Delta n/n)_C$ for the disappearance of the 1/2 FQHS were only $\simeq 20 \%$, and this lack of robustness to charge distribution asymmetry in fact led Suen \textit{et al.} to conclude that the 1/2 FQHS had a 2C origin \cite{Suen.PRL.1994}. In contrast, our data in Fig. \hyperlink{fig2}{3(c)} show that the 1/2 FQHS is stable even for $(\Delta n/n)_C$ exceeding $30 \%$, consistent with its having a 1C origin. We emphasize that our conclusion is based not only on the absolute value of $(\Delta n/n)_C$, but also on its trend as a function of \textit{n}, namely that the 1/2 FQHS is much more robust to charge distribution asymmetry when the 2DES exhibits 1C features.

\begin{figure}[t!]
\hypertarget{fig2}{}
\includegraphics[width=1\columnwidth]{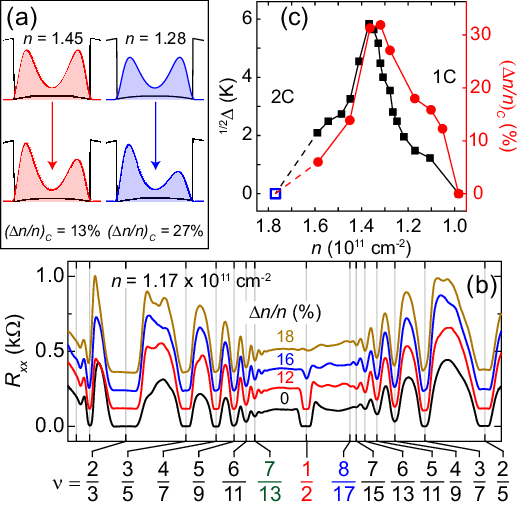}
\caption{\label{fig:fig2}(a) Calculated charge distributions at \textit{n} = 1.45 (red) and \textit{n} = 1.28 (blue) in a symmetric and asymmetric QW. (b) Characteristic $R_{xx}$ traces at \textit{n} = 1.17 showing the evolution of the 1/2 FQHS as the charge distribution is made asymmetric. (c) The critical layer density imbalance $(\Delta n/n)_C$ required to destabilize the 1/2 FQHS, plotted together with $^{1/2}\Delta$, as a function of \textit{n}. The 1/2 FQHSs on the high-density side of the $^{1/2}\Delta$ peak have smaller $(\Delta n/n)_C$.}
\end{figure}

Several theory groups have focused on the competition between the 1C (Pfaffian) and 2C ($\Psi_{331}$) FQHSs at $\nu=1/2$ in wide QWs \cite{Nomura.JPSJ.2004, Peterson.PRB.2010, Zhu.PRB.2016}. Their calculations, which ignore the CF Fermi sea as a competing state, generally indicate a transition between a 1C FQHS at large $\alpha$ to a 2C FQHS at small $\alpha$. In Fig. \hyperlink{fig3}{2(b)} we show the results of Refs. \cite{Peterson.PRB.2010, Zhu.PRB.2016}; {also see notes EM5 and EM6}. Similar to the experimental data, the theoretical results exhibit a non-monotonic dependence of $^{1/2}\Delta$ on $\alpha$ with a maximum energy at an intermediate $\alpha$. At first sight, the similarity of the theoretical and experimental results is quite impressive in that $\alpha$ at which the gap reaches its maximum value, as well as the magnitude of the maximum gap, are in reasonably good agreement. 

However, there are several noteworthy caveats. First, it is surprising that the measured gaps near their maximum exceed the calculated values. Measured gaps for FQHSs are generally smaller than the gaps calculated for ideal 2DESs because of finite (non-zero) electron layer thickness, LL mixing, and disorder in real samples \cite{Willett.PRB.1988, Melik-Alaverdian.PRB.1995, Park.JPhys.1999, Morf.PRB.2002, Storni.PRL.2010, VillegasRosales.PRL.2021}. The role of finite layer thickness is presumably included in Refs. \cite{Peterson.PRB.2010, Zhu.PRB.2016} as these studies consider charge distributions that are appropriate for 2DESs in wide QWs. But, to make the calculations feasible, they use only ``model" charge distributions, and it is possible that the calculated energy gaps are inaccurate as they can sensitively depend on the exact charge distributions (see SM Sections IV and V \cite{SM.2025} for a more detailed discussion).

Second, compared to the theoretical curves, the experimental data in Fig. \hyperlink{fig3}{2(b)} exhibit a much sharper peak and a much narrower range of $\alpha$ where the 1/2 FQHS is stable. This might reflect the fact that the energy gap calculations in Refs. \cite{Peterson.PRB.2010, Zhu.PRB.2016} do not consider the competition with the CF Fermi sea (for large values of $\alpha$) or a bilayer Wigner crystal (for small $\alpha$). This shortcoming of the calculations is clear considering that both the experimental data [Fig. \hyperlink{fig3}{2(b)}] and recent theoretical results \cite{Sharma.PRB.2024} show that in our wide QW the 1/2 FQHS gives way to a CF Fermi sea at $\alpha > 0.11$ while the calculations in Refs. \cite{Peterson.PRB.2010, Zhu.PRB.2016} predict that the 1C FQHS continues to persist up to much larger $\alpha$.

{Another important feature of Fig. \hyperlink{fig3}{2} data is the critical $\alpha$ ($\alpha_C$) at which the transition between the 1C and 2C 1/2 FQHS occurs. Our experimental data [Fig. \hyperlink{fig3}{2(a)}] indicate that the 2DES undergoes its 1C to 2C transition at $\alpha_C \simeq 0.064$, which is just to the left of the observed peak in $^{1/2}\Delta$ [Fig. \hyperlink{fig3}{2(b)}]. We note that $\alpha_C \simeq 0.064$ is obtained from the 1C to 2C transitions observed at other fillings near 1/2. References \cite{Peterson.PRB.2010, Zhu.PRB.2016} predict that the 1C to 2C transition of the 1/2 FQHS occurs in the vicinity of the peak $^{1/2}\Delta$; see SM Section VI \cite{SM.2025}. Regardless of the precise value of $\alpha_C$, it is remarkable that both experimental data and calculations indicate that the 1/2 FQHS energy gap does not collapse when the 2DES makes a 1C to 2C transition, and in fact shows a \textit{maximum} near the transition, with a persisting Hall plateau {(see note EM7)}.} This is surprising, although it has been suggested that the 1C Pfaffian state can continuously evolve to the 2C $\Psi_{331}$ state \cite{Ho.PRL.1995, Halperin.Surf.Sci.1994, Read.PRB.2000, Wen.PRL.2000} by varying the interlayer tunneling. References \cite{Nomura.JPSJ.2004, Zhu.PRB.2016} identify that the peak in $^{1/2}\Delta$ likely occurs because of a level crossing at $\nu = 1/2$ between the lowest excited state and higher energy levels with 1C and 2C character. {This is also discussed by Wen \cite{Wen.PRL.2000} in his theory of continuous topological phase transitions, characteristics of which include finite charge gap but collapse of the neutral gap. In the special case of the Pfaffian and $\Psi_{331}$ states, a continuous topological phase transition involves a neutral Majorana mode at the critical point, while the charged excitations remain gapped \cite{Wen.PRL.2000, Yang.PRB.2017}.} {To summarize, our data provide clear evidence for a 1C to 2C transition of the 2DES on the flanks of $\nu=1/2$, and suggest that the 1/2 FQHS also makes such a transition {(see note EM8)}. A topological phase transition between a 1C (non-Abelian) and 2C (Abelian) FQHSs, to the best of our knowledge, is unique to the 1/2 FQHS in wide GaAs QWs, opening avenues to experiments in quantum criticality such as those proposed in Ref. \cite{Ma.PRB.2022}.}

We close by remarking that even-denominator FQHSs at half-filled LLs have been reported in the excited $N = 1$ LL in several platforms such as 2DESs in GaAs \cite{Willett.PRL.1987, Pan.PRL.1999, Chung.NatMater.2021}, ZnO \cite{Falson.Nat.Phys.2015}, AlAs \cite{Shafayat.PRL.2018}, monolayer, bilayer and trilayer graphene \cite{Ki.NanoLett.2014, Review.Dean.Kim.Li.Young.2020, Kim.NatPhys.2019, Huang.PRX.2022, Assouline.PRL.2024, Kumar.Nat.Comm.2025, Hu.Nat.Phys.2025, Chanda.Preprint.2024}, and WSe\textsubscript{2} \cite{Shi.NatureNanotech.2020}. They have also been reported in the $N=0$ LL in 2D hole systems confined to wide GaAs QWs \cite{Liu.PRL.2014, Liu.PRB.2014}, and 2DESs in wide AlAs QWs \cite{Shafayat.PRL.2023}. Very recently, 2D hole systems in ultrahigh-quality GaAs QWs have also revealed a number of even-denominator FQHSs at unusual fillings, e.g. at $\nu=3/4$, 3/8, 3/10, and 1/4 \cite{Wang.PRL.2022, Wang.PNAS.2023, Wang.PRL.2023}. We would like to emphasize that the 6 K gap we measure for the 1/2 FQHS in our sample is the largest transport gap ever reported for any even-denominator FQHS. This is particularly impressive considering that our sample size ($\simeq 20$ mm$^2$) is macroscopic, and is about $10^7$ larger than the typical size of bilayer graphene samples where comparable energy gaps (up to $\simeq 5$ K) have been reported \cite{Assouline.PRL.2024}. The very large energy gap and sample size, together with the evidence that the 1/2 FQHS we observe is 1C, most likely a Pfaffian (non-Abelian) state, make this FQHS a prime candidate platform for topological quantum computing. 

\begin{acknowledgments}

We acknowledge support by the National Science Foundation (NSF) Grant No. DMR 2104771 for measurements, and the Gordon and Betty Moore Foundation’s EPiQS Initiative (Grant No. GBMF9615 to L.N.P.) for sample fabrication. A portion of this work was performed at the National High Magnetic Field Laboratory, which is supported by National Science Foundation Cooperative Agreement No. DMR-2128556 and the State of Florida. This research is funded in part by QuantEmX travel grants from ICAM and the Gordon and Betty Moore Foundation through Grant GBMF9616 to S.K.S., C.W. and A.G. We thank R. Nowell, G. Jones, A. Bangura and T. Murphy at NHMFL for technical assistance, and J. K. Jain, D. Sheng, F. D. M. Haldane, M. Peterson, N. Regnault and S. Das Sarma for illuminating discussions. 

\end{acknowledgments}

\bibliography{aps}

\begin{thebibliography}{14}%
\makeatletter
\providecommand \@ifxundefined [1]{%
 \@ifx{#1\undefined}
}%
\providecommand \@ifnum [1]{%
 \ifnum #1\expandafter \@firstoftwo
 \else \expandafter \@secondoftwo
 \fi
}%
\providecommand \@ifx [1]{%
 \ifx #1\expandafter \@firstoftwo
 \else \expandafter \@secondoftwo
 \fi
}%
\providecommand \natexlab [1]{#1}%
\providecommand \enquote  [1]{``#1''}%
\providecommand \bibnamefont  [1]{#1}%
\providecommand \bibfnamefont [1]{#1}%
\providecommand \citenamefont [1]{#1}%
\providecommand \href@noop [0]{\@secondoftwo}%
\providecommand \href [0]{\begingroup \@sanitize@url \@href}%
\providecommand \@href[1]{\@@startlink{#1}\@@href}%
\providecommand \@@href[1]{\endgroup#1\@@endlink}%
\providecommand \@sanitize@url [0]{\catcode `\\12\catcode `\$12\catcode `\&12\catcode `\#12\catcode `\^12\catcode `\_12\catcode `\%12\relax}%
\providecommand \@@startlink[1]{}%
\providecommand \@@endlink[0]{}%
\providecommand \url  [0]{\begingroup\@sanitize@url \@url }%
\providecommand \@url [1]{\endgroup\@href {#1}{\urlprefix }}%
\providecommand \urlprefix  [0]{URL }%
\providecommand \Eprint [0]{\href }%
\providecommand \doibase [0]{https://doi.org/}%
\providecommand \selectlanguage [0]{\@gobble}%
\providecommand \bibinfo  [0]{\@secondoftwo}%
\providecommand \bibfield  [0]{\@secondoftwo}%
\providecommand \translation [1]{[#1]}%
\providecommand \BibitemOpen [0]{}%
\providecommand \bibitemStop [0]{}%
\providecommand \bibitemNoStop [0]{.\EOS\space}%
\providecommand \EOS [0]{\spacefactor3000\relax}%
\providecommand \BibitemShut  [1]{\csname bibitem#1\endcsname}%
\let\auto@bib@innerbib\@empty
\bibitem [{\citenamefont {Suen}\ \emph {et~al.}(1992{\natexlab{a}})\citenamefont {Suen}, \citenamefont {Engel}, \citenamefont {Santos}, \citenamefont {Shayegan},\ and\ \citenamefont {Tsui}}]{Suen.PRL.1992}%
  \BibitemOpen
  \bibfield  {author} {\bibinfo {author} {\bibfnamefont {Y.~W.}\ \bibnamefont {Suen}}, \bibinfo {author} {\bibfnamefont {L.~W.}\ \bibnamefont {Engel}}, \bibinfo {author} {\bibfnamefont {M.~B.}\ \bibnamefont {Santos}}, \bibinfo {author} {\bibfnamefont {M.}~\bibnamefont {Shayegan}},\ and\ \bibinfo {author} {\bibfnamefont {D.~C.}\ \bibnamefont {Tsui}},\ }\bibfield  {title} {\bibinfo {title} {{Observation of a \ensuremath{\nu} = 1/2 fractional quantum Hall state in a double-layer electron system}},\ }\href {https://doi.org/10.1103/PhysRevLett.68.1379} {\bibfield  {journal} {\bibinfo  {journal} {Phys. Rev. Lett.}\ }\textbf {\bibinfo {volume} {68}},\ \bibinfo {pages} {1379} (\bibinfo {year} {1992}{\natexlab{a}})}\BibitemShut {NoStop}%
\bibitem [{\citenamefont {Suen}\ \emph {et~al.}(1992{\natexlab{b}})\citenamefont {Suen}, \citenamefont {Santos},\ and\ \citenamefont {Shayegan}}]{Suen2.PRL.1992}%
  \BibitemOpen
  \bibfield  {author} {\bibinfo {author} {\bibfnamefont {Y.~W.}\ \bibnamefont {Suen}}, \bibinfo {author} {\bibfnamefont {M.~B.}\ \bibnamefont {Santos}},\ and\ \bibinfo {author} {\bibfnamefont {M.}~\bibnamefont {Shayegan}},\ }\bibfield  {title} {\bibinfo {title} {{Correlated states of an electron system in a wide quantum well}},\ }\href {https://doi.org/10.1103/PhysRevLett.69.3551} {\bibfield  {journal} {\bibinfo  {journal} {Phys. Rev. Lett.}\ }\textbf {\bibinfo {volume} {69}},\ \bibinfo {pages} {3551} (\bibinfo {year} {1992}{\natexlab{b}})}\BibitemShut {NoStop}%
\bibitem [{\citenamefont {Suen}\ \emph {et~al.}(1994)\citenamefont {Suen}, \citenamefont {Manoharan}, \citenamefont {Ying}, \citenamefont {Santos},\ and\ \citenamefont {Shayegan}}]{Suen.PRL.1994}%
  \BibitemOpen
  \bibfield  {author} {\bibinfo {author} {\bibfnamefont {Y.~W.}\ \bibnamefont {Suen}}, \bibinfo {author} {\bibfnamefont {H.~C.}\ \bibnamefont {Manoharan}}, \bibinfo {author} {\bibfnamefont {X.}~\bibnamefont {Ying}}, \bibinfo {author} {\bibfnamefont {M.~B.}\ \bibnamefont {Santos}},\ and\ \bibinfo {author} {\bibfnamefont {M.}~\bibnamefont {Shayegan}},\ }\bibfield  {title} {\bibinfo {title} {{Origin of the \ensuremath{\nu} = 1/2 fractional quantum Hall state in wide single quantum wells}},\ }\href {https://doi.org/10.1103/PhysRevLett.72.3405} {\bibfield  {journal} {\bibinfo  {journal} {Phys. Rev. Lett.}\ }\textbf {\bibinfo {volume} {72}},\ \bibinfo {pages} {3405} (\bibinfo {year} {1994})}\BibitemShut {NoStop}%
\bibitem [{\citenamefont {Manoharan}\ \emph {et~al.}(1996)\citenamefont {Manoharan}, \citenamefont {Suen}, \citenamefont {Santos},\ and\ \citenamefont {Shayegan}}]{Manoharan.PRL.1996}%
  \BibitemOpen
  \bibfield  {author} {\bibinfo {author} {\bibfnamefont {H.~C.}\ \bibnamefont {Manoharan}}, \bibinfo {author} {\bibfnamefont {Y.~W.}\ \bibnamefont {Suen}}, \bibinfo {author} {\bibfnamefont {M.~B.}\ \bibnamefont {Santos}},\ and\ \bibinfo {author} {\bibfnamefont {M.}~\bibnamefont {Shayegan}},\ }\bibfield  {title} {\bibinfo {title} {{Evidence for a Bilayer Quantum Wigner Solid}},\ }\href {https://doi.org/10.1103/PhysRevLett.77.1813} {\bibfield  {journal} {\bibinfo  {journal} {Phys. Rev. Lett.}\ }\textbf {\bibinfo {volume} {77}},\ \bibinfo {pages} {1813} (\bibinfo {year} {1996})}\BibitemShut {NoStop}%
\bibitem [{\citenamefont {Shayegan}(1999)}]{Shayegan.Review.LesHouches.1999}%
  \BibitemOpen
  \bibfield  {author} {\bibinfo {author} {\bibfnamefont {M.}~\bibnamefont {Shayegan}},\ }\bibfield  {title} {\bibinfo {title} {{Electrons in a Flatland}},\ }in\ \href {https://doi.org/https://doi.org/10.1007/3-540-46637-1_1} {\emph {\bibinfo {booktitle} {{1998 Les Houches Summer School, Session LXIX, Topological Aspects of Low Dimensional Systems}}}},\ \bibinfo {series and number} {NATO Advanced Study Institute},\ \bibinfo {editor} {edited by\ \bibinfo {editor} {\bibfnamefont {A.}~\bibnamefont {Comtet}}, \bibinfo {editor} {\bibfnamefont {T.}~\bibnamefont {Jolic{\oe}ur}}, \bibinfo {editor} {\bibfnamefont {S.}~\bibnamefont {Ouvry}},\ and\ \bibinfo {editor} {\bibfnamefont {F.}~\bibnamefont {David}}}\ (\bibinfo {organization} {Springer-Verlag},\ \bibinfo {address} {Berlin},\ \bibinfo {year} {1999})\ pp.\ \bibinfo {pages} {1--51}\BibitemShut {NoStop}%
\bibitem [{\citenamefont {Chung}\ \emph {et~al.}(2021)\citenamefont {Chung}, \citenamefont {Villegas~Rosales}, \citenamefont {Baldwin}, \citenamefont {Madathil}, \citenamefont {West}, \citenamefont {Shayegan},\ and\ \citenamefont {Pfeiffer}}]{Chung.NatMater.2021}%
  \BibitemOpen
  \bibfield  {author} {\bibinfo {author} {\bibfnamefont {Y.~J.}\ \bibnamefont {Chung}}, \bibinfo {author} {\bibfnamefont {K.}~\bibnamefont {Villegas~Rosales}}, \bibinfo {author} {\bibfnamefont {K.}~\bibnamefont {Baldwin}}, \bibinfo {author} {\bibfnamefont {P.}~\bibnamefont {Madathil}}, \bibinfo {author} {\bibfnamefont {K.}~\bibnamefont {West}}, \bibinfo {author} {\bibfnamefont {M.}~\bibnamefont {Shayegan}},\ and\ \bibinfo {author} {\bibfnamefont {L.}~\bibnamefont {Pfeiffer}},\ }\bibfield  {title} {\bibinfo {title} {{Ultra-high-quality two-dimensional electron systems}},\ }\href {https://doi.org/10.1038/s41563-021-00942-3} {\bibfield  {journal} {\bibinfo  {journal} {Nat. Mater.}\ }\textbf {\bibinfo {volume} {20}},\ \bibinfo {pages} {632} (\bibinfo {year} {2021})}\BibitemShut {NoStop}%
\bibitem [{\citenamefont {Assouline}\ \emph {et~al.}(2024)\citenamefont {Assouline}, \citenamefont {Wang}, \citenamefont {Zhou}, \citenamefont {Cohen}, \citenamefont {Yang}, \citenamefont {Zhang}, \citenamefont {Taniguchi}, \citenamefont {Watanabe}, \citenamefont {Mong}, \citenamefont {Zaletel},\ and\ \citenamefont {Young}}]{Assouline.PRL.2024}%
  \BibitemOpen
  \bibfield  {author} {\bibinfo {author} {\bibfnamefont {A.}~\bibnamefont {Assouline}}, \bibinfo {author} {\bibfnamefont {T.}~\bibnamefont {Wang}}, \bibinfo {author} {\bibfnamefont {H.}~\bibnamefont {Zhou}}, \bibinfo {author} {\bibfnamefont {L.~A.}\ \bibnamefont {Cohen}}, \bibinfo {author} {\bibfnamefont {F.}~\bibnamefont {Yang}}, \bibinfo {author} {\bibfnamefont {R.}~\bibnamefont {Zhang}}, \bibinfo {author} {\bibfnamefont {T.}~\bibnamefont {Taniguchi}}, \bibinfo {author} {\bibfnamefont {K.}~\bibnamefont {Watanabe}}, \bibinfo {author} {\bibfnamefont {R.~S.~K.}\ \bibnamefont {Mong}}, \bibinfo {author} {\bibfnamefont {M.~P.}\ \bibnamefont {Zaletel}},\ and\ \bibinfo {author} {\bibfnamefont {A.~F.}\ \bibnamefont {Young}},\ }\bibfield  {title} {\bibinfo {title} {{Energy Gap of the Even-Denominator Fractional Quantum Hall State in Bilayer Graphene}},\ }\href {https://doi.org/10.1103/PhysRevLett.132.046603} {\bibfield  {journal} {\bibinfo  {journal} {Phys. Rev. Lett.}\ }\textbf {\bibinfo {volume} {132}},\ \bibinfo
  {pages} {046603} (\bibinfo {year} {2024})}\BibitemShut {NoStop}%
\bibitem [{\citenamefont {Shabani}\ \emph {et~al.}(2013)\citenamefont {Shabani}, \citenamefont {Liu}, \citenamefont {Shayegan}, \citenamefont {Pfeiffer}, \citenamefont {West},\ and\ \citenamefont {Baldwin}}]{Shabani.PRB.2013}%
  \BibitemOpen
  \bibfield  {author} {\bibinfo {author} {\bibfnamefont {J.}~\bibnamefont {Shabani}}, \bibinfo {author} {\bibfnamefont {Y.}~\bibnamefont {Liu}}, \bibinfo {author} {\bibfnamefont {M.}~\bibnamefont {Shayegan}}, \bibinfo {author} {\bibfnamefont {L.~N.}\ \bibnamefont {Pfeiffer}}, \bibinfo {author} {\bibfnamefont {K.~W.}\ \bibnamefont {West}},\ and\ \bibinfo {author} {\bibfnamefont {K.~W.}\ \bibnamefont {Baldwin}},\ }\bibfield  {title} {\bibinfo {title} {{Phase diagrams for the stability of the $\ensuremath{\nu}=\frac{1}{2}$ fractional quantum Hall effect in electron systems confined to symmetric, wide GaAs quantum wells}},\ }\href {https://doi.org/10.1103/PhysRevB.88.245413} {\bibfield  {journal} {\bibinfo  {journal} {Phys. Rev. B}\ }\textbf {\bibinfo {volume} {88}},\ \bibinfo {pages} {245413} (\bibinfo {year} {2013})}\BibitemShut {NoStop}%
\bibitem [{\citenamefont {Peterson}\ and\ \citenamefont {Das~Sarma}(2010)}]{Peterson.PRB.2010}%
  \BibitemOpen
  \bibfield  {author} {\bibinfo {author} {\bibfnamefont {M.~R.}\ \bibnamefont {Peterson}}\ and\ \bibinfo {author} {\bibfnamefont {S.}~\bibnamefont {Das~Sarma}},\ }\bibfield  {title} {\bibinfo {title} {{Quantum Hall phase diagram of half-filled bilayers in the lowest and the second orbital Landau levels: Abelian versus non-Abelian incompressible fractional quantum Hall states}},\ }\href {https://doi.org/10.1103/PhysRevB.81.165304} {\bibfield  {journal} {\bibinfo  {journal} {Phys. Rev. B}\ }\textbf {\bibinfo {volume} {81}},\ \bibinfo {pages} {165304} (\bibinfo {year} {2010})}\BibitemShut {NoStop}%
\bibitem [{\citenamefont {Thiebaut}\ \emph {et~al.}(2015)\citenamefont {Thiebaut}, \citenamefont {Regnault},\ and\ \citenamefont {Goerbig}}]{Thiebaut.PRB.2015}%
  \BibitemOpen
  \bibfield  {author} {\bibinfo {author} {\bibfnamefont {N.}~\bibnamefont {Thiebaut}}, \bibinfo {author} {\bibfnamefont {N.}~\bibnamefont {Regnault}},\ and\ \bibinfo {author} {\bibfnamefont {M.~O.}\ \bibnamefont {Goerbig}},\ }\bibfield  {title} {\bibinfo {title} {{Fractional quantum Hall states versus Wigner crystals in wide quantum wells in the half-filled lowest and second Landau levels}},\ }\href {https://doi.org/10.1103/PhysRevB.92.245401} {\bibfield  {journal} {\bibinfo  {journal} {Phys. Rev. B}\ }\textbf {\bibinfo {volume} {92}},\ \bibinfo {pages} {245401} (\bibinfo {year} {2015})}\BibitemShut {NoStop}%
\bibitem [{\citenamefont {Zhu}\ \emph {et~al.}(2016)\citenamefont {Zhu}, \citenamefont {Liu}, \citenamefont {Haldane},\ and\ \citenamefont {Sheng}}]{Zhu.PRB.2016}%
  \BibitemOpen
  \bibfield  {author} {\bibinfo {author} {\bibfnamefont {W.}~\bibnamefont {Zhu}}, \bibinfo {author} {\bibfnamefont {Z.}~\bibnamefont {Liu}}, \bibinfo {author} {\bibfnamefont {F.~D.~M.}\ \bibnamefont {Haldane}},\ and\ \bibinfo {author} {\bibfnamefont {D.~N.}\ \bibnamefont {Sheng}},\ }\bibfield  {title} {\bibinfo {title} {{Fractional quantum Hall bilayers at half filling: Tunneling-driven non-Abelian phase}},\ }\href {https://doi.org/10.1103/PhysRevB.94.245147} {\bibfield  {journal} {\bibinfo  {journal} {Phys. Rev. B}\ }\textbf {\bibinfo {volume} {94}},\ \bibinfo {pages} {245147} (\bibinfo {year} {2016})}\BibitemShut {NoStop}%
\bibitem [{\citenamefont {Sharma}\ \emph {et~al.}(2024)\citenamefont {Sharma}, \citenamefont {Balram},\ and\ \citenamefont {Jain}}]{Sharma.PRB.2024}%
  \BibitemOpen
  \bibfield  {author} {\bibinfo {author} {\bibfnamefont {A.}~\bibnamefont {Sharma}}, \bibinfo {author} {\bibfnamefont {A.~C.}\ \bibnamefont {Balram}},\ and\ \bibinfo {author} {\bibfnamefont {J.~K.}\ \bibnamefont {Jain}},\ }\bibfield  {title} {\bibinfo {title} {{Composite-fermion pairing at half-filled and quarter-filled lowest Landau level}},\ }\href {https://doi.org/10.1103/PhysRevB.109.035306} {\bibfield  {journal} {\bibinfo  {journal} {Phys. Rev. B}\ }\textbf {\bibinfo {volume} {109}},\ \bibinfo {pages} {035306} (\bibinfo {year} {2024})}\BibitemShut {NoStop}%
\bibitem [{\citenamefont {Villegas~Rosales}\ \emph {et~al.}(2021)\citenamefont {Villegas~Rosales}, \citenamefont {Madathil}, \citenamefont {Chung}, \citenamefont {Pfeiffer}, \citenamefont {West}, \citenamefont {Baldwin},\ and\ \citenamefont {Shayegan}}]{VillegasRosales.PRL.2021}%
  \BibitemOpen
  \bibfield  {author} {\bibinfo {author} {\bibfnamefont {K.~A.}\ \bibnamefont {Villegas~Rosales}}, \bibinfo {author} {\bibfnamefont {P.~T.}\ \bibnamefont {Madathil}}, \bibinfo {author} {\bibfnamefont {Y.~J.}\ \bibnamefont {Chung}}, \bibinfo {author} {\bibfnamefont {L.~N.}\ \bibnamefont {Pfeiffer}}, \bibinfo {author} {\bibfnamefont {K.~W.}\ \bibnamefont {West}}, \bibinfo {author} {\bibfnamefont {K.~W.}\ \bibnamefont {Baldwin}},\ and\ \bibinfo {author} {\bibfnamefont {M.}~\bibnamefont {Shayegan}},\ }\bibfield  {title} {\bibinfo {title} {{Fractional Quantum Hall Effect Energy Gaps: Role of Electron Layer Thickness}},\ }\href {https://doi.org/10.1103/PhysRevLett.127.056801} {\bibfield  {journal} {\bibinfo  {journal} {Phys. Rev. Lett.}\ }\textbf {\bibinfo {volume} {127}},\ \bibinfo {pages} {056801} (\bibinfo {year} {2021})}\BibitemShut {NoStop}%
\bibitem [{\citenamefont {Singh}\ \emph {et~al.}(2024)\citenamefont {Singh}, \citenamefont {Wang}, \citenamefont {Tai}, \citenamefont {Calhoun}, \citenamefont {Villegas~Rosales}, \citenamefont {Madathil}, \citenamefont {Gupta}, \citenamefont {Baldwin}, \citenamefont {Pfeiffer},\ and\ \citenamefont {Shayegan}}]{Singh.NatPhys.2024}%
  \BibitemOpen
  \bibfield  {author} {\bibinfo {author} {\bibfnamefont {S.~K.}\ \bibnamefont {Singh}}, \bibinfo {author} {\bibfnamefont {C.}~\bibnamefont {Wang}}, \bibinfo {author} {\bibfnamefont {C.~T.}\ \bibnamefont {Tai}}, \bibinfo {author} {\bibfnamefont {C.~S.}\ \bibnamefont {Calhoun}}, \bibinfo {author} {\bibfnamefont {K.~A.}\ \bibnamefont {Villegas~Rosales}}, \bibinfo {author} {\bibfnamefont {P.~T.}\ \bibnamefont {Madathil}}, \bibinfo {author} {\bibfnamefont {A.}~\bibnamefont {Gupta}}, \bibinfo {author} {\bibfnamefont {K.~W.}\ \bibnamefont {Baldwin}}, \bibinfo {author} {\bibfnamefont {L.~N.}\ \bibnamefont {Pfeiffer}},\ and\ \bibinfo {author} {\bibfnamefont {M.}~\bibnamefont {Shayegan}},\ }\bibfield  {title} {\bibinfo {title} {{Topological phase transition between Jain states and daughter states of the $\nu$ = 1/2 fractional quantum Hall state}},\ }\href {https://doi.org/10.1038/s41567-024-02517-w} {\bibfield  {journal} {\bibinfo  {journal} {Nat. Phys.}\ }\textbf {\bibinfo {volume} {20}},\ \bibinfo {pages} {1247}
  (\bibinfo {year} {2024})}\BibitemShut {NoStop}%
\end{thebibliography}%


\begin{thebibliography}{72}%
\makeatletter
\providecommand \@ifxundefined [1]{%
 \@ifx{#1\undefined}
}%
\providecommand \@ifnum [1]{%
 \ifnum #1\expandafter \@firstoftwo
 \else \expandafter \@secondoftwo
 \fi
}%
\providecommand \@ifx [1]{%
 \ifx #1\expandafter \@firstoftwo
 \else \expandafter \@secondoftwo
 \fi
}%
\providecommand \natexlab [1]{#1}%
\providecommand \enquote  [1]{``#1''}%
\providecommand \bibnamefont  [1]{#1}%
\providecommand \bibfnamefont [1]{#1}%
\providecommand \citenamefont [1]{#1}%
\providecommand \href@noop [0]{\@secondoftwo}%
\providecommand \href [0]{\begingroup \@sanitize@url \@href}%
\providecommand \@href[1]{\@@startlink{#1}\@@href}%
\providecommand \@@href[1]{\endgroup#1\@@endlink}%
\providecommand \@sanitize@url [0]{\catcode `\\12\catcode `\$12\catcode `\&12\catcode `\#12\catcode `\^12\catcode `\_12\catcode `\%12\relax}%
\providecommand \@@startlink[1]{}%
\providecommand \@@endlink[0]{}%
\providecommand \url  [0]{\begingroup\@sanitize@url \@url }%
\providecommand \@url [1]{\endgroup\@href {#1}{\urlprefix }}%
\providecommand \urlprefix  [0]{URL }%
\providecommand \Eprint [0]{\href }%
\providecommand \doibase [0]{https://doi.org/}%
\providecommand \selectlanguage [0]{\@gobble}%
\providecommand \bibinfo  [0]{\@secondoftwo}%
\providecommand \bibfield  [0]{\@secondoftwo}%
\providecommand \translation [1]{[#1]}%
\providecommand \BibitemOpen [0]{}%
\providecommand \bibitemStop [0]{}%
\providecommand \bibitemNoStop [0]{.\EOS\space}%
\providecommand \EOS [0]{\spacefactor3000\relax}%
\providecommand \BibitemShut  [1]{\csname bibitem#1\endcsname}%
\let\auto@bib@innerbib\@empty
\bibitem [{\citenamefont {Willett}\ \emph {et~al.}(1987)\citenamefont {Willett}, \citenamefont {Eisenstein}, \citenamefont {St\"ormer}, \citenamefont {Tsui}, \citenamefont {Gossard},\ and\ \citenamefont {English}}]{Willett.PRL.1987}%
  \BibitemOpen
  \bibfield  {author} {\bibinfo {author} {\bibfnamefont {R.}~\bibnamefont {Willett}}, \bibinfo {author} {\bibfnamefont {J.~P.}\ \bibnamefont {Eisenstein}}, \bibinfo {author} {\bibfnamefont {H.~L.}\ \bibnamefont {St\"ormer}}, \bibinfo {author} {\bibfnamefont {D.~C.}\ \bibnamefont {Tsui}}, \bibinfo {author} {\bibfnamefont {A.~C.}\ \bibnamefont {Gossard}},\ and\ \bibinfo {author} {\bibfnamefont {J.~H.}\ \bibnamefont {English}},\ }\bibfield  {title} {\bibinfo {title} {{Observation of an even-denominator quantum number in the fractional quantum Hall effect}},\ }\href {https://doi.org/10.1103/PhysRevLett.59.1776} {\bibfield  {journal} {\bibinfo  {journal} {Phys. Rev. Lett.}\ }\textbf {\bibinfo {volume} {59}},\ \bibinfo {pages} {1776} (\bibinfo {year} {1987})}\BibitemShut {NoStop}%
\bibitem [{\citenamefont {Nayak}\ \emph {et~al.}(2008)\citenamefont {Nayak}, \citenamefont {Simon}, \citenamefont {Stern}, \citenamefont {Freedman},\ and\ \citenamefont {Das~Sarma}}]{Nayak.RevModPhys.2008}%
  \BibitemOpen
  \bibfield  {author} {\bibinfo {author} {\bibfnamefont {C.}~\bibnamefont {Nayak}}, \bibinfo {author} {\bibfnamefont {S.~H.}\ \bibnamefont {Simon}}, \bibinfo {author} {\bibfnamefont {A.}~\bibnamefont {Stern}}, \bibinfo {author} {\bibfnamefont {M.}~\bibnamefont {Freedman}},\ and\ \bibinfo {author} {\bibfnamefont {S.}~\bibnamefont {Das~Sarma}},\ }\bibfield  {title} {\bibinfo {title} {{Non-Abelian anyons and topological quantum computation}},\ }\href {https://doi.org/10.1103/RevModPhys.80.1083} {\bibfield  {journal} {\bibinfo  {journal} {Rev. Mod. Phys.}\ }\textbf {\bibinfo {volume} {80}},\ \bibinfo {pages} {1083} (\bibinfo {year} {2008})}\BibitemShut {NoStop}%
\bibitem [{\citenamefont {Banerjee}\ \emph {et~al.}(2018)\citenamefont {Banerjee}, \citenamefont {Heiblum}, \citenamefont {Umansky}, \citenamefont {Feldman}, \citenamefont {Oreg},\ and\ \citenamefont {Stern}}]{Banerjee.Nature.2018}%
  \BibitemOpen
  \bibfield  {author} {\bibinfo {author} {\bibfnamefont {M.}~\bibnamefont {Banerjee}}, \bibinfo {author} {\bibfnamefont {M.}~\bibnamefont {Heiblum}}, \bibinfo {author} {\bibfnamefont {V.}~\bibnamefont {Umansky}}, \bibinfo {author} {\bibfnamefont {D.~E.}\ \bibnamefont {Feldman}}, \bibinfo {author} {\bibfnamefont {Y.}~\bibnamefont {Oreg}},\ and\ \bibinfo {author} {\bibfnamefont {A.}~\bibnamefont {Stern}},\ }\bibfield  {title} {\bibinfo {title} {{Observation of half-integer thermal Hall conductance}},\ }\href {https://doi.org/10.1038/s41586-018-0184-1} {\bibfield  {journal} {\bibinfo  {journal} {Nature}\ }\textbf {\bibinfo {volume} {559}},\ \bibinfo {pages} {205} (\bibinfo {year} {2018})}\BibitemShut {NoStop}%
\bibitem [{\citenamefont {Willett}\ \emph {et~al.}(2023)\citenamefont {Willett}, \citenamefont {Shtengel}, \citenamefont {Nayak}, \citenamefont {Pfeiffer}, \citenamefont {Chung}, \citenamefont {Peabody}, \citenamefont {Baldwin},\ and\ \citenamefont {West}}]{Willett.PRX.2023}%
  \BibitemOpen
  \bibfield  {author} {\bibinfo {author} {\bibfnamefont {R.~L.}\ \bibnamefont {Willett}}, \bibinfo {author} {\bibfnamefont {K.}~\bibnamefont {Shtengel}}, \bibinfo {author} {\bibfnamefont {C.}~\bibnamefont {Nayak}}, \bibinfo {author} {\bibfnamefont {L.~N.}\ \bibnamefont {Pfeiffer}}, \bibinfo {author} {\bibfnamefont {Y.~J.}\ \bibnamefont {Chung}}, \bibinfo {author} {\bibfnamefont {M.~L.}\ \bibnamefont {Peabody}}, \bibinfo {author} {\bibfnamefont {K.~W.}\ \bibnamefont {Baldwin}},\ and\ \bibinfo {author} {\bibfnamefont {K.~W.}\ \bibnamefont {West}},\ }\bibfield  {title} {\bibinfo {title} {{Interference Measurements of Non-Abelian $e/4$ $\&$ Abelian $e/2$ Quasiparticle Braiding}},\ }\href {https://doi.org/10.1103/PhysRevX.13.011028} {\bibfield  {journal} {\bibinfo  {journal} {Phys. Rev. X}\ }\textbf {\bibinfo {volume} {13}},\ \bibinfo {pages} {011028} (\bibinfo {year} {2023})}\BibitemShut {NoStop}%
\bibitem [{\citenamefont {Moore}\ and\ \citenamefont {Read}(1991)}]{MooreRead.NuclPhysB.1990}%
  \BibitemOpen
  \bibfield  {author} {\bibinfo {author} {\bibfnamefont {G.}~\bibnamefont {Moore}}\ and\ \bibinfo {author} {\bibfnamefont {N.}~\bibnamefont {Read}},\ }\bibfield  {title} {\bibinfo {title} {{Nonabelions in the fractional quantum Hall effect}},\ }\href {https://doi.org/https://doi.org/10.1016/0550-3213(91)90407-O} {\bibfield  {journal} {\bibinfo  {journal} {Nucl. Phys. B}\ }\textbf {\bibinfo {volume} {360}},\ \bibinfo {pages} {362} (\bibinfo {year} {1991})}\BibitemShut {NoStop}%
\bibitem [{\citenamefont {Lee}\ \emph {et~al.}(2007)\citenamefont {Lee}, \citenamefont {Ryu}, \citenamefont {Nayak},\ and\ \citenamefont {Fisher}}]{Lee.PRL.2007}%
  \BibitemOpen
  \bibfield  {author} {\bibinfo {author} {\bibfnamefont {S.-S.}\ \bibnamefont {Lee}}, \bibinfo {author} {\bibfnamefont {S.}~\bibnamefont {Ryu}}, \bibinfo {author} {\bibfnamefont {C.}~\bibnamefont {Nayak}},\ and\ \bibinfo {author} {\bibfnamefont {M.~P.~A.}\ \bibnamefont {Fisher}},\ }\bibfield  {title} {\bibinfo {title} {{Particle-Hole Symmetry and the $\ensuremath{\nu}=\frac{5}{2}$ Quantum Hall State}},\ }\href {https://doi.org/10.1103/PhysRevLett.99.236807} {\bibfield  {journal} {\bibinfo  {journal} {Phys. Rev. Lett.}\ }\textbf {\bibinfo {volume} {99}},\ \bibinfo {pages} {236807} (\bibinfo {year} {2007})}\BibitemShut {NoStop}%
\bibitem [{\citenamefont {Levin}\ \emph {et~al.}(2007)\citenamefont {Levin}, \citenamefont {Halperin},\ and\ \citenamefont {Rosenow}}]{Levin.PRL.2007}%
  \BibitemOpen
  \bibfield  {author} {\bibinfo {author} {\bibfnamefont {M.}~\bibnamefont {Levin}}, \bibinfo {author} {\bibfnamefont {B.~I.}\ \bibnamefont {Halperin}},\ and\ \bibinfo {author} {\bibfnamefont {B.}~\bibnamefont {Rosenow}},\ }\bibfield  {title} {\bibinfo {title} {{Particle-Hole Symmetry and the Pfaffian State}},\ }\href {https://doi.org/10.1103/PhysRevLett.99.236806} {\bibfield  {journal} {\bibinfo  {journal} {Phys. Rev. Lett.}\ }\textbf {\bibinfo {volume} {99}},\ \bibinfo {pages} {236806} (\bibinfo {year} {2007})}\BibitemShut {NoStop}%
\bibitem [{\citenamefont {Son}(2015)}]{Son.PRX.2015}%
  \BibitemOpen
  \bibfield  {author} {\bibinfo {author} {\bibfnamefont {D.~T.}\ \bibnamefont {Son}},\ }\bibfield  {title} {\bibinfo {title} {{Is the Composite Fermion a Dirac Particle?}},\ }\href {https://doi.org/10.1103/PhysRevX.5.031027} {\bibfield  {journal} {\bibinfo  {journal} {Phys. Rev. X}\ }\textbf {\bibinfo {volume} {5}},\ \bibinfo {pages} {031027} (\bibinfo {year} {2015})}\BibitemShut {NoStop}%
\bibitem [{\citenamefont {Scarola}\ \emph {et~al.}(2000)\citenamefont {Scarola}, \citenamefont {Park},\ and\ \citenamefont {Jain}}]{Scarola.Nature.2000}%
  \BibitemOpen
  \bibfield  {author} {\bibinfo {author} {\bibfnamefont {V.~W.}\ \bibnamefont {Scarola}}, \bibinfo {author} {\bibfnamefont {K.}~\bibnamefont {Park}},\ and\ \bibinfo {author} {\bibfnamefont {J.~K.}\ \bibnamefont {Jain}},\ }\bibfield  {title} {\bibinfo {title} {{Cooper instability of composite fermions}},\ }\href {https://doi.org/10.1038/35022524} {\bibfield  {journal} {\bibinfo  {journal} {Nature}\ }\textbf {\bibinfo {volume} {406}},\ \bibinfo {pages} {863} (\bibinfo {year} {2000})}\BibitemShut {NoStop}%
\bibitem [{\citenamefont {Das~Sarma}\ \emph {et~al.}(2005)\citenamefont {Das~Sarma}, \citenamefont {Freedman},\ and\ \citenamefont {Nayak}}]{Sarma.PRL.2005}%
  \BibitemOpen
  \bibfield  {author} {\bibinfo {author} {\bibfnamefont {S.}~\bibnamefont {Das~Sarma}}, \bibinfo {author} {\bibfnamefont {M.}~\bibnamefont {Freedman}},\ and\ \bibinfo {author} {\bibfnamefont {C.}~\bibnamefont {Nayak}},\ }\bibfield  {title} {\bibinfo {title} {{Topologically Protected Qubits from a Possible Non-Abelian Fractional Quantum Hall State}},\ }\href {https://doi.org/10.1103/PhysRevLett.94.166802} {\bibfield  {journal} {\bibinfo  {journal} {Phys. Rev. Lett.}\ }\textbf {\bibinfo {volume} {94}},\ \bibinfo {pages} {166802} (\bibinfo {year} {2005})}\BibitemShut {NoStop}%
\bibitem [{\citenamefont {Suen}\ \emph {et~al.}(1992{\natexlab{a}})\citenamefont {Suen}, \citenamefont {Engel}, \citenamefont {Santos}, \citenamefont {Shayegan},\ and\ \citenamefont {Tsui}}]{Suen.PRL.1992}%
  \BibitemOpen
  \bibfield  {author} {\bibinfo {author} {\bibfnamefont {Y.~W.}\ \bibnamefont {Suen}}, \bibinfo {author} {\bibfnamefont {L.~W.}\ \bibnamefont {Engel}}, \bibinfo {author} {\bibfnamefont {M.~B.}\ \bibnamefont {Santos}}, \bibinfo {author} {\bibfnamefont {M.}~\bibnamefont {Shayegan}},\ and\ \bibinfo {author} {\bibfnamefont {D.~C.}\ \bibnamefont {Tsui}},\ }\bibfield  {title} {\bibinfo {title} {{Observation of a \ensuremath{\nu} = 1/2 fractional quantum Hall state in a double-layer electron system}},\ }\href {https://doi.org/10.1103/PhysRevLett.68.1379} {\bibfield  {journal} {\bibinfo  {journal} {Phys. Rev. Lett.}\ }\textbf {\bibinfo {volume} {68}},\ \bibinfo {pages} {1379} (\bibinfo {year} {1992}{\natexlab{a}})}\BibitemShut {NoStop}%
\bibitem [{\citenamefont {Suen}\ \emph {et~al.}(1992{\natexlab{b}})\citenamefont {Suen}, \citenamefont {Santos},\ and\ \citenamefont {Shayegan}}]{Suen2.PRL.1992}%
  \BibitemOpen
  \bibfield  {author} {\bibinfo {author} {\bibfnamefont {Y.~W.}\ \bibnamefont {Suen}}, \bibinfo {author} {\bibfnamefont {M.~B.}\ \bibnamefont {Santos}},\ and\ \bibinfo {author} {\bibfnamefont {M.}~\bibnamefont {Shayegan}},\ }\bibfield  {title} {\bibinfo {title} {{Correlated states of an electron system in a wide quantum well}},\ }\href {https://doi.org/10.1103/PhysRevLett.69.3551} {\bibfield  {journal} {\bibinfo  {journal} {Phys. Rev. Lett.}\ }\textbf {\bibinfo {volume} {69}},\ \bibinfo {pages} {3551} (\bibinfo {year} {1992}{\natexlab{b}})}\BibitemShut {NoStop}%
\bibitem [{\citenamefont {Suen}\ \emph {et~al.}(1994)\citenamefont {Suen}, \citenamefont {Manoharan}, \citenamefont {Ying}, \citenamefont {Santos},\ and\ \citenamefont {Shayegan}}]{Suen.PRL.1994}%
  \BibitemOpen
  \bibfield  {author} {\bibinfo {author} {\bibfnamefont {Y.~W.}\ \bibnamefont {Suen}}, \bibinfo {author} {\bibfnamefont {H.~C.}\ \bibnamefont {Manoharan}}, \bibinfo {author} {\bibfnamefont {X.}~\bibnamefont {Ying}}, \bibinfo {author} {\bibfnamefont {M.~B.}\ \bibnamefont {Santos}},\ and\ \bibinfo {author} {\bibfnamefont {M.}~\bibnamefont {Shayegan}},\ }\bibfield  {title} {\bibinfo {title} {{Origin of the \ensuremath{\nu} = 1/2 fractional quantum Hall state in wide single quantum wells}},\ }\href {https://doi.org/10.1103/PhysRevLett.72.3405} {\bibfield  {journal} {\bibinfo  {journal} {Phys. Rev. Lett.}\ }\textbf {\bibinfo {volume} {72}},\ \bibinfo {pages} {3405} (\bibinfo {year} {1994})}\BibitemShut {NoStop}%
\bibitem [{\citenamefont {Eisenstein}\ \emph {et~al.}(1992)\citenamefont {Eisenstein}, \citenamefont {Boebinger}, \citenamefont {Pfeiffer}, \citenamefont {West},\ and\ \citenamefont {He}}]{Eisenstein.PRL.1992}%
  \BibitemOpen
  \bibfield  {author} {\bibinfo {author} {\bibfnamefont {J.~P.}\ \bibnamefont {Eisenstein}}, \bibinfo {author} {\bibfnamefont {G.~S.}\ \bibnamefont {Boebinger}}, \bibinfo {author} {\bibfnamefont {L.~N.}\ \bibnamefont {Pfeiffer}}, \bibinfo {author} {\bibfnamefont {K.~W.}\ \bibnamefont {West}},\ and\ \bibinfo {author} {\bibfnamefont {S.}~\bibnamefont {He}},\ }\bibfield  {title} {\bibinfo {title} {{New fractional quantum Hall state in double-layer two-dimensional electron systems}},\ }\href {https://doi.org/10.1103/PhysRevLett.68.1383} {\bibfield  {journal} {\bibinfo  {journal} {Phys. Rev. Lett.}\ }\textbf {\bibinfo {volume} {68}},\ \bibinfo {pages} {1383} (\bibinfo {year} {1992})}\BibitemShut {NoStop}%
\bibitem [{\citenamefont {Suen}\ \emph {et~al.}(1991)\citenamefont {Suen}, \citenamefont {Jo}, \citenamefont {Santos}, \citenamefont {Engel}, \citenamefont {Hwang},\ and\ \citenamefont {Shayegan}}]{Suen.PRB.1991}%
  \BibitemOpen
  \bibfield  {author} {\bibinfo {author} {\bibfnamefont {Y.~W.}\ \bibnamefont {Suen}}, \bibinfo {author} {\bibfnamefont {J.}~\bibnamefont {Jo}}, \bibinfo {author} {\bibfnamefont {M.~B.}\ \bibnamefont {Santos}}, \bibinfo {author} {\bibfnamefont {L.~W.}\ \bibnamefont {Engel}}, \bibinfo {author} {\bibfnamefont {S.~W.}\ \bibnamefont {Hwang}},\ and\ \bibinfo {author} {\bibfnamefont {M.}~\bibnamefont {Shayegan}},\ }\bibfield  {title} {\bibinfo {title} {{Missing integral quantum Hall effect in a wide single quantum well}},\ }\href {https://doi.org/10.1103/PhysRevB.44.5947} {\bibfield  {journal} {\bibinfo  {journal} {Phys. Rev. B}\ }\textbf {\bibinfo {volume} {44}},\ \bibinfo {pages} {5947} (\bibinfo {year} {1991})}\BibitemShut {NoStop}%
\bibitem [{\citenamefont {Shabani}\ \emph {et~al.}(2009{\natexlab{a}})\citenamefont {Shabani}, \citenamefont {Gokmen},\ and\ \citenamefont {Shayegan}}]{Shabani.PRL.2009}%
  \BibitemOpen
  \bibfield  {author} {\bibinfo {author} {\bibfnamefont {J.}~\bibnamefont {Shabani}}, \bibinfo {author} {\bibfnamefont {T.}~\bibnamefont {Gokmen}},\ and\ \bibinfo {author} {\bibfnamefont {M.}~\bibnamefont {Shayegan}},\ }\bibfield  {title} {\bibinfo {title} {{Correlated States of Electrons in Wide Quantum Wells at Low Fillings: The Role of Charge Distribution Symmetry}},\ }\href {https://doi.org/10.1103/PhysRevLett.103.046805} {\bibfield  {journal} {\bibinfo  {journal} {Phys. Rev. Lett.}\ }\textbf {\bibinfo {volume} {103}},\ \bibinfo {pages} {046805} (\bibinfo {year} {2009}{\natexlab{a}})}\BibitemShut {NoStop}%
\bibitem [{\citenamefont {Shabani}\ \emph {et~al.}(2009{\natexlab{b}})\citenamefont {Shabani}, \citenamefont {Gokmen}, \citenamefont {Chiu},\ and\ \citenamefont {Shayegan}}]{Shabani2.PRL.2009}%
  \BibitemOpen
  \bibfield  {author} {\bibinfo {author} {\bibfnamefont {J.}~\bibnamefont {Shabani}}, \bibinfo {author} {\bibfnamefont {T.}~\bibnamefont {Gokmen}}, \bibinfo {author} {\bibfnamefont {Y.~T.}\ \bibnamefont {Chiu}},\ and\ \bibinfo {author} {\bibfnamefont {M.}~\bibnamefont {Shayegan}},\ }\bibfield  {title} {\bibinfo {title} {{Evidence for Developing Fractional Quantum Hall States at Even Denominator $1/2$ and $1/4$ Fillings in Asymmetric Wide Quantum Wells}},\ }\href {https://doi.org/10.1103/PhysRevLett.103.256802} {\bibfield  {journal} {\bibinfo  {journal} {Phys. Rev. Lett.}\ }\textbf {\bibinfo {volume} {103}},\ \bibinfo {pages} {256802} (\bibinfo {year} {2009}{\natexlab{b}})}\BibitemShut {NoStop}%
\bibitem [{\citenamefont {Shabani}\ \emph {et~al.}(2013)\citenamefont {Shabani}, \citenamefont {Liu}, \citenamefont {Shayegan}, \citenamefont {Pfeiffer}, \citenamefont {West},\ and\ \citenamefont {Baldwin}}]{Shabani.PRB.2013}%
  \BibitemOpen
  \bibfield  {author} {\bibinfo {author} {\bibfnamefont {J.}~\bibnamefont {Shabani}}, \bibinfo {author} {\bibfnamefont {Y.}~\bibnamefont {Liu}}, \bibinfo {author} {\bibfnamefont {M.}~\bibnamefont {Shayegan}}, \bibinfo {author} {\bibfnamefont {L.~N.}\ \bibnamefont {Pfeiffer}}, \bibinfo {author} {\bibfnamefont {K.~W.}\ \bibnamefont {West}},\ and\ \bibinfo {author} {\bibfnamefont {K.~W.}\ \bibnamefont {Baldwin}},\ }\bibfield  {title} {\bibinfo {title} {{Phase diagrams for the stability of the $\ensuremath{\nu}=\frac{1}{2}$ fractional quantum Hall effect in electron systems confined to symmetric, wide GaAs quantum wells}},\ }\href {https://doi.org/10.1103/PhysRevB.88.245413} {\bibfield  {journal} {\bibinfo  {journal} {Phys. Rev. B}\ }\textbf {\bibinfo {volume} {88}},\ \bibinfo {pages} {245413} (\bibinfo {year} {2013})}\BibitemShut {NoStop}%
\bibitem [{\citenamefont {Mueed}\ \emph {et~al.}(2015)\citenamefont {Mueed}, \citenamefont {Kamburov}, \citenamefont {Hasdemir}, \citenamefont {Shayegan}, \citenamefont {Pfeiffer}, \citenamefont {West},\ and\ \citenamefont {Baldwin}}]{Mueed.PRL.2015}%
  \BibitemOpen
  \bibfield  {author} {\bibinfo {author} {\bibfnamefont {M.~A.}\ \bibnamefont {Mueed}}, \bibinfo {author} {\bibfnamefont {D.}~\bibnamefont {Kamburov}}, \bibinfo {author} {\bibfnamefont {S.}~\bibnamefont {Hasdemir}}, \bibinfo {author} {\bibfnamefont {M.}~\bibnamefont {Shayegan}}, \bibinfo {author} {\bibfnamefont {L.~N.}\ \bibnamefont {Pfeiffer}}, \bibinfo {author} {\bibfnamefont {K.~W.}\ \bibnamefont {West}},\ and\ \bibinfo {author} {\bibfnamefont {K.~W.}\ \bibnamefont {Baldwin}},\ }\bibfield  {title} {\bibinfo {title} {{Geometric Resonance of Composite Fermions Near the $\ensuremath{\nu}=1/2$ Fractional Quantum Hall State}},\ }\href {https://doi.org/10.1103/PhysRevLett.114.236406} {\bibfield  {journal} {\bibinfo  {journal} {Phys. Rev. Lett.}\ }\textbf {\bibinfo {volume} {114}},\ \bibinfo {pages} {236406} (\bibinfo {year} {2015})}\BibitemShut {NoStop}%
\bibitem [{\citenamefont {Mueed}\ \emph {et~al.}(2016)\citenamefont {Mueed}, \citenamefont {Kamburov}, \citenamefont {Pfeiffer}, \citenamefont {West}, \citenamefont {Baldwin},\ and\ \citenamefont {Shayegan}}]{Mueed.PRL.2016}%
  \BibitemOpen
  \bibfield  {author} {\bibinfo {author} {\bibfnamefont {M.~A.}\ \bibnamefont {Mueed}}, \bibinfo {author} {\bibfnamefont {D.}~\bibnamefont {Kamburov}}, \bibinfo {author} {\bibfnamefont {L.~N.}\ \bibnamefont {Pfeiffer}}, \bibinfo {author} {\bibfnamefont {K.~W.}\ \bibnamefont {West}}, \bibinfo {author} {\bibfnamefont {K.~W.}\ \bibnamefont {Baldwin}},\ and\ \bibinfo {author} {\bibfnamefont {M.}~\bibnamefont {Shayegan}},\ }\bibfield  {title} {\bibinfo {title} {{Geometric Resonance of Composite Fermions near Bilayer Quantum Hall States}},\ }\href {https://doi.org/10.1103/PhysRevLett.117.246801} {\bibfield  {journal} {\bibinfo  {journal} {Phys. Rev. Lett.}\ }\textbf {\bibinfo {volume} {117}},\ \bibinfo {pages} {246801} (\bibinfo {year} {2016})}\BibitemShut {NoStop}%
\bibitem [{\citenamefont {Hasdemir}\ \emph {et~al.}(2015)\citenamefont {Hasdemir}, \citenamefont {Liu}, \citenamefont {Deng}, \citenamefont {Shayegan}, \citenamefont {Pfeiffer}, \citenamefont {West}, \citenamefont {Baldwin},\ and\ \citenamefont {Winkler}}]{Hasdemir.PRB.2017}%
  \BibitemOpen
  \bibfield  {author} {\bibinfo {author} {\bibfnamefont {S.}~\bibnamefont {Hasdemir}}, \bibinfo {author} {\bibfnamefont {Y.}~\bibnamefont {Liu}}, \bibinfo {author} {\bibfnamefont {H.}~\bibnamefont {Deng}}, \bibinfo {author} {\bibfnamefont {M.}~\bibnamefont {Shayegan}}, \bibinfo {author} {\bibfnamefont {L.~N.}\ \bibnamefont {Pfeiffer}}, \bibinfo {author} {\bibfnamefont {K.~W.}\ \bibnamefont {West}}, \bibinfo {author} {\bibfnamefont {K.~W.}\ \bibnamefont {Baldwin}},\ and\ \bibinfo {author} {\bibfnamefont {R.}~\bibnamefont {Winkler}},\ }\bibfield  {title} {\bibinfo {title} {{$\ensuremath{\nu}=1/2$ fractional quantum Hall effect in tilted magnetic fields}},\ }\href {https://doi.org/10.1103/PhysRevB.91.045113} {\bibfield  {journal} {\bibinfo  {journal} {Phys. Rev. B}\ }\textbf {\bibinfo {volume} {91}},\ \bibinfo {pages} {045113} (\bibinfo {year} {2015})}\BibitemShut {NoStop}%
\bibitem [{\citenamefont {Singh}\ \emph {et~al.}(2024)\citenamefont {Singh}, \citenamefont {Wang}, \citenamefont {Tai}, \citenamefont {Calhoun}, \citenamefont {Villegas~Rosales}, \citenamefont {Madathil}, \citenamefont {Gupta}, \citenamefont {Baldwin}, \citenamefont {Pfeiffer},\ and\ \citenamefont {Shayegan}}]{Singh.NatPhys.2024}%
  \BibitemOpen
  \bibfield  {author} {\bibinfo {author} {\bibfnamefont {S.~K.}\ \bibnamefont {Singh}}, \bibinfo {author} {\bibfnamefont {C.}~\bibnamefont {Wang}}, \bibinfo {author} {\bibfnamefont {C.~T.}\ \bibnamefont {Tai}}, \bibinfo {author} {\bibfnamefont {C.~S.}\ \bibnamefont {Calhoun}}, \bibinfo {author} {\bibfnamefont {K.~A.}\ \bibnamefont {Villegas~Rosales}}, \bibinfo {author} {\bibfnamefont {P.~T.}\ \bibnamefont {Madathil}}, \bibinfo {author} {\bibfnamefont {A.}~\bibnamefont {Gupta}}, \bibinfo {author} {\bibfnamefont {K.~W.}\ \bibnamefont {Baldwin}}, \bibinfo {author} {\bibfnamefont {L.~N.}\ \bibnamefont {Pfeiffer}},\ and\ \bibinfo {author} {\bibfnamefont {M.}~\bibnamefont {Shayegan}},\ }\bibfield  {title} {\bibinfo {title} {{Topological phase transition between Jain states and daughter states of the $\nu$ = 1/2 fractional quantum Hall state}},\ }\href {https://doi.org/10.1038/s41567-024-02517-w} {\bibfield  {journal} {\bibinfo  {journal} {Nat. Phys.}\ }\textbf {\bibinfo {volume} {20}},\ \bibinfo {pages} {1247}
  (\bibinfo {year} {2024})}\BibitemShut {NoStop}%
\bibitem [{\citenamefont {Greiter}\ \emph {et~al.}(1992{\natexlab{a}})\citenamefont {Greiter}, \citenamefont {Wen},\ and\ \citenamefont {Wilczek}}]{Greiter.PRB.1992}%
  \BibitemOpen
  \bibfield  {author} {\bibinfo {author} {\bibfnamefont {M.}~\bibnamefont {Greiter}}, \bibinfo {author} {\bibfnamefont {X.~G.}\ \bibnamefont {Wen}},\ and\ \bibinfo {author} {\bibfnamefont {F.}~\bibnamefont {Wilczek}},\ }\bibfield  {title} {\bibinfo {title} {{Paired Hall states in double-layer electron systems}},\ }\href {https://doi.org/10.1103/PhysRevB.46.9586} {\bibfield  {journal} {\bibinfo  {journal} {Phys. Rev. B}\ }\textbf {\bibinfo {volume} {46}},\ \bibinfo {pages} {9586} (\bibinfo {year} {1992}{\natexlab{a}})}\BibitemShut {NoStop}%
\bibitem [{\citenamefont {Greiter}\ \emph {et~al.}(1992{\natexlab{b}})\citenamefont {Greiter}, \citenamefont {Wen},\ and\ \citenamefont {Wilczek}}]{Greiter.NucPhysB.1992}%
  \BibitemOpen
  \bibfield  {author} {\bibinfo {author} {\bibfnamefont {M.}~\bibnamefont {Greiter}}, \bibinfo {author} {\bibfnamefont {X.-G.}\ \bibnamefont {Wen}},\ and\ \bibinfo {author} {\bibfnamefont {F.}~\bibnamefont {Wilczek}},\ }\bibfield  {title} {\bibinfo {title} {{Paired Hall states}},\ }\href {https://doi.org/10.1016/0550-3213(92)90401-V} {\bibfield  {journal} {\bibinfo  {journal} {Nucl. Phys. B}\ }\textbf {\bibinfo {volume} {374}},\ \bibinfo {pages} {567} (\bibinfo {year} {1992}{\natexlab{b}})}\BibitemShut {NoStop}%
\bibitem [{\citenamefont {He}\ \emph {et~al.}(1993)\citenamefont {He}, \citenamefont {Das~Sarma},\ and\ \citenamefont {Xie}}]{He.PRB.1993}%
  \BibitemOpen
  \bibfield  {author} {\bibinfo {author} {\bibfnamefont {S.}~\bibnamefont {He}}, \bibinfo {author} {\bibfnamefont {S.}~\bibnamefont {Das~Sarma}},\ and\ \bibinfo {author} {\bibfnamefont {X.~C.}\ \bibnamefont {Xie}},\ }\bibfield  {title} {\bibinfo {title} {{Quantized Hall effect and quantum phase transitions in coupled two-layer electron systems}},\ }\href {https://doi.org/10.1103/PhysRevB.47.4394} {\bibfield  {journal} {\bibinfo  {journal} {Phys. Rev. B}\ }\textbf {\bibinfo {volume} {47}},\ \bibinfo {pages} {4394} (\bibinfo {year} {1993})}\BibitemShut {NoStop}%
\bibitem [{\citenamefont {Nomura}\ and\ \citenamefont {Yoshioka}(2004)}]{Nomura.JPSJ.2004}%
  \BibitemOpen
  \bibfield  {author} {\bibinfo {author} {\bibfnamefont {K.}~\bibnamefont {Nomura}}\ and\ \bibinfo {author} {\bibfnamefont {D.}~\bibnamefont {Yoshioka}},\ }\bibfield  {title} {\bibinfo {title} {{Gap Evolution in $\nu = 1/2$ Bilayer Quantum Hall Systems}},\ }\href {https://doi.org/10.1143/jpsj.73.2612} {\bibfield  {journal} {\bibinfo  {journal} {J. Phys. Soc. Jpn.}\ }\textbf {\bibinfo {volume} {73}},\ \bibinfo {pages} {2612} (\bibinfo {year} {2004})}\BibitemShut {NoStop}%
\bibitem [{\citenamefont {Peterson}\ and\ \citenamefont {Das~Sarma}(2010)}]{Peterson.PRB.2010}%
  \BibitemOpen
  \bibfield  {author} {\bibinfo {author} {\bibfnamefont {M.~R.}\ \bibnamefont {Peterson}}\ and\ \bibinfo {author} {\bibfnamefont {S.}~\bibnamefont {Das~Sarma}},\ }\bibfield  {title} {\bibinfo {title} {{Quantum Hall phase diagram of half-filled bilayers in the lowest and the second orbital Landau levels: Abelian versus non-Abelian incompressible fractional quantum Hall states}},\ }\href {https://doi.org/10.1103/PhysRevB.81.165304} {\bibfield  {journal} {\bibinfo  {journal} {Phys. Rev. B}\ }\textbf {\bibinfo {volume} {81}},\ \bibinfo {pages} {165304} (\bibinfo {year} {2010})}\BibitemShut {NoStop}%
\bibitem [{\citenamefont {Thiebaut}\ \emph {et~al.}(2015)\citenamefont {Thiebaut}, \citenamefont {Regnault},\ and\ \citenamefont {Goerbig}}]{Thiebaut.PRB.2015}%
  \BibitemOpen
  \bibfield  {author} {\bibinfo {author} {\bibfnamefont {N.}~\bibnamefont {Thiebaut}}, \bibinfo {author} {\bibfnamefont {N.}~\bibnamefont {Regnault}},\ and\ \bibinfo {author} {\bibfnamefont {M.~O.}\ \bibnamefont {Goerbig}},\ }\bibfield  {title} {\bibinfo {title} {{Fractional quantum Hall states versus Wigner crystals in wide quantum wells in the half-filled lowest and second Landau levels}},\ }\href {https://doi.org/10.1103/PhysRevB.92.245401} {\bibfield  {journal} {\bibinfo  {journal} {Phys. Rev. B}\ }\textbf {\bibinfo {volume} {92}},\ \bibinfo {pages} {245401} (\bibinfo {year} {2015})}\BibitemShut {NoStop}%
\bibitem [{\citenamefont {Zhu}\ \emph {et~al.}(2016)\citenamefont {Zhu}, \citenamefont {Liu}, \citenamefont {Haldane},\ and\ \citenamefont {Sheng}}]{Zhu.PRB.2016}%
  \BibitemOpen
  \bibfield  {author} {\bibinfo {author} {\bibfnamefont {W.}~\bibnamefont {Zhu}}, \bibinfo {author} {\bibfnamefont {Z.}~\bibnamefont {Liu}}, \bibinfo {author} {\bibfnamefont {F.~D.~M.}\ \bibnamefont {Haldane}},\ and\ \bibinfo {author} {\bibfnamefont {D.~N.}\ \bibnamefont {Sheng}},\ }\bibfield  {title} {\bibinfo {title} {{Fractional quantum Hall bilayers at half filling: Tunneling-driven non-Abelian phase}},\ }\href {https://doi.org/10.1103/PhysRevB.94.245147} {\bibfield  {journal} {\bibinfo  {journal} {Phys. Rev. B}\ }\textbf {\bibinfo {volume} {94}},\ \bibinfo {pages} {245147} (\bibinfo {year} {2016})}\BibitemShut {NoStop}%
\bibitem [{\citenamefont {Sharma}\ \emph {et~al.}(2024)\citenamefont {Sharma}, \citenamefont {Balram},\ and\ \citenamefont {Jain}}]{Sharma.PRB.2024}%
  \BibitemOpen
  \bibfield  {author} {\bibinfo {author} {\bibfnamefont {A.}~\bibnamefont {Sharma}}, \bibinfo {author} {\bibfnamefont {A.~C.}\ \bibnamefont {Balram}},\ and\ \bibinfo {author} {\bibfnamefont {J.~K.}\ \bibnamefont {Jain}},\ }\bibfield  {title} {\bibinfo {title} {{Composite-fermion pairing at half-filled and quarter-filled lowest Landau level}},\ }\href {https://doi.org/10.1103/PhysRevB.109.035306} {\bibfield  {journal} {\bibinfo  {journal} {Phys. Rev. B}\ }\textbf {\bibinfo {volume} {109}},\ \bibinfo {pages} {035306} (\bibinfo {year} {2024})}\BibitemShut {NoStop}%
\bibitem [{\citenamefont {Halperin}(1983)}]{Halperin.Helv.Phys.Acta.1983}%
  \BibitemOpen
  \bibfield  {author} {\bibinfo {author} {\bibfnamefont {B.~I.}\ \bibnamefont {Halperin}},\ }\bibfield  {title} {\bibinfo {title} {{Theory of the quantized Hall conductance}},\ }\href@noop {} {\bibfield  {journal} {\bibinfo  {journal} {Helv. Phys. Acta}\ }\textbf {\bibinfo {volume} {56}},\ \bibinfo {pages} {75} (\bibinfo {year} {1983})}\BibitemShut {NoStop}%
\bibitem [{\citenamefont {Levin}\ and\ \citenamefont {Halperin}(2009)}]{Levin.PRB.2009}%
  \BibitemOpen
  \bibfield  {author} {\bibinfo {author} {\bibfnamefont {M.}~\bibnamefont {Levin}}\ and\ \bibinfo {author} {\bibfnamefont {B.~I.}\ \bibnamefont {Halperin}},\ }\bibfield  {title} {\bibinfo {title} {{Collective states of non-Abelian quasiparticles in a magnetic field}},\ }\href {https://doi.org/10.1103/PhysRevB.79.205301} {\bibfield  {journal} {\bibinfo  {journal} {Phys. Rev. B}\ }\textbf {\bibinfo {volume} {79}},\ \bibinfo {pages} {205301} (\bibinfo {year} {2009})}\BibitemShut {NoStop}%
\bibitem [{\citenamefont {Yutushui}\ \emph {et~al.}(2024)\citenamefont {Yutushui}, \citenamefont {Hermanns},\ and\ \citenamefont {Mross}}]{Yutushui.PRB.2024}%
  \BibitemOpen
  \bibfield  {author} {\bibinfo {author} {\bibfnamefont {M.}~\bibnamefont {Yutushui}}, \bibinfo {author} {\bibfnamefont {M.}~\bibnamefont {Hermanns}},\ and\ \bibinfo {author} {\bibfnamefont {D.~F.}\ \bibnamefont {Mross}},\ }\bibfield  {title} {\bibinfo {title} {{Paired fermions in strong magnetic fields and daughters of even-denominator Hall plateaus}},\ }\href {https://doi.org/10.1103/PhysRevB.110.165402} {\bibfield  {journal} {\bibinfo  {journal} {Phys. Rev. B}\ }\textbf {\bibinfo {volume} {110}},\ \bibinfo {pages} {165402} (\bibinfo {year} {2024})}\BibitemShut {NoStop}%
\bibitem [{\citenamefont {Zheltonozhskii}\ \emph {et~al.}(2024)\citenamefont {Zheltonozhskii}, \citenamefont {Stern},\ and\ \citenamefont {Lindner}}]{Zheltonozhskii.PRB.2024}%
  \BibitemOpen
  \bibfield  {author} {\bibinfo {author} {\bibfnamefont {E.}~\bibnamefont {Zheltonozhskii}}, \bibinfo {author} {\bibfnamefont {A.}~\bibnamefont {Stern}},\ and\ \bibinfo {author} {\bibfnamefont {N.~H.}\ \bibnamefont {Lindner}},\ }\bibfield  {title} {\bibinfo {title} {{Identifying the topological order of quantized half-filled Landau levels through their daughter states}},\ }\href {https://doi.org/10.1103/PhysRevB.110.245140} {\bibfield  {journal} {\bibinfo  {journal} {Phys. Rev. B}\ }\textbf {\bibinfo {volume} {110}},\ \bibinfo {pages} {245140} (\bibinfo {year} {2024})}\BibitemShut {NoStop}%
\bibitem [{\citenamefont {Manoharan}\ \emph {et~al.}(1996)\citenamefont {Manoharan}, \citenamefont {Suen}, \citenamefont {Santos},\ and\ \citenamefont {Shayegan}}]{Manoharan.PRL.1996}%
  \BibitemOpen
  \bibfield  {author} {\bibinfo {author} {\bibfnamefont {H.~C.}\ \bibnamefont {Manoharan}}, \bibinfo {author} {\bibfnamefont {Y.~W.}\ \bibnamefont {Suen}}, \bibinfo {author} {\bibfnamefont {M.~B.}\ \bibnamefont {Santos}},\ and\ \bibinfo {author} {\bibfnamefont {M.}~\bibnamefont {Shayegan}},\ }\bibfield  {title} {\bibinfo {title} {{Evidence for a Bilayer Quantum Wigner Solid}},\ }\href {https://doi.org/10.1103/PhysRevLett.77.1813} {\bibfield  {journal} {\bibinfo  {journal} {Phys. Rev. Lett.}\ }\textbf {\bibinfo {volume} {77}},\ \bibinfo {pages} {1813} (\bibinfo {year} {1996})}\BibitemShut {NoStop}%
\bibitem [{\citenamefont {Manoharan}\ \emph {et~al.}(1997)\citenamefont {Manoharan}, \citenamefont {Suen}, \citenamefont {Lay}, \citenamefont {Santos},\ and\ \citenamefont {Shayegan}}]{Manoharan.PRL.1997}%
  \BibitemOpen
  \bibfield  {author} {\bibinfo {author} {\bibfnamefont {H.~C.}\ \bibnamefont {Manoharan}}, \bibinfo {author} {\bibfnamefont {Y.~W.}\ \bibnamefont {Suen}}, \bibinfo {author} {\bibfnamefont {T.~S.}\ \bibnamefont {Lay}}, \bibinfo {author} {\bibfnamefont {M.~B.}\ \bibnamefont {Santos}},\ and\ \bibinfo {author} {\bibfnamefont {M.}~\bibnamefont {Shayegan}},\ }\bibfield  {title} {\bibinfo {title} {{Spontaneous Interlayer Charge Transfer near the Magnetic Quantum Limit}},\ }\href {https://doi.org/10.1103/PhysRevLett.79.2722} {\bibfield  {journal} {\bibinfo  {journal} {Phys. Rev. Lett.}\ }\textbf {\bibinfo {volume} {79}},\ \bibinfo {pages} {2722} (\bibinfo {year} {1997})}\BibitemShut {NoStop}%
\bibitem [{\citenamefont {Shayegan}(1999)}]{Shayegan.Review.LesHouches.1999}%
  \BibitemOpen
  \bibfield  {author} {\bibinfo {author} {\bibfnamefont {M.}~\bibnamefont {Shayegan}},\ }\bibfield  {title} {\bibinfo {title} {{Electrons in a Flatland}},\ }in\ \href {https://doi.org/https://doi.org/10.1007/3-540-46637-1_1} {\emph {\bibinfo {booktitle} {{1998 Les Houches Summer School, Session LXIX, Topological Aspects of Low Dimensional Systems}}}},\ \bibinfo {series and number} {NATO Advanced Study Institute},\ \bibinfo {editor} {edited by\ \bibinfo {editor} {\bibfnamefont {A.}~\bibnamefont {Comtet}}, \bibinfo {editor} {\bibfnamefont {T.}~\bibnamefont {Jolic{\oe}ur}}, \bibinfo {editor} {\bibfnamefont {S.}~\bibnamefont {Ouvry}},\ and\ \bibinfo {editor} {\bibfnamefont {F.}~\bibnamefont {David}}}\ (\bibinfo {organization} {Springer-Verlag},\ \bibinfo {address} {Berlin},\ \bibinfo {year} {1999})\ pp.\ \bibinfo {pages} {1--51}\BibitemShut {NoStop}%
\bibitem [{SM.()}]{SM.2025}%
  \BibitemOpen
  \href@noop {} {}\bibinfo {note} {{See Supplemental Material for an exhaustive set of $R_{xx}$ traces, Arrhenius plots for the extraction of $^{1/2}\Delta$ and relevant discussion around the physical parameters of the 2DES in a wide GaAs QW.}}\BibitemShut {Stop}%
\bibitem [{\citenamefont {Jain}(2007)}]{Jain.composite.fermions.2007}%
  \BibitemOpen
  \bibfield  {author} {\bibinfo {author} {\bibfnamefont {J.~K.}\ \bibnamefont {Jain}},\ }\href {https://doi.org/https://doi.org/10.1017/CBO9780511607561} {\emph {\bibinfo {title} {{Composite Fermions}}}}\ (\bibinfo  {publisher} {Cambridge University Press},\ \bibinfo {address} {Cambridge, England},\ \bibinfo {year} {2007})\BibitemShut {NoStop}%
\bibitem [{\citenamefont {Hatke}\ \emph {et~al.}(2017)\citenamefont {Hatke}, \citenamefont {Liu}, \citenamefont {Engel}, \citenamefont {Pfeiffer}, \citenamefont {West}, \citenamefont {Baldwin},\ and\ \citenamefont {Shayegan}}]{Hatke.PRB.2017}%
  \BibitemOpen
  \bibfield  {author} {\bibinfo {author} {\bibfnamefont {A.~T.}\ \bibnamefont {Hatke}}, \bibinfo {author} {\bibfnamefont {Y.}~\bibnamefont {Liu}}, \bibinfo {author} {\bibfnamefont {L.~W.}\ \bibnamefont {Engel}}, \bibinfo {author} {\bibfnamefont {L.~N.}\ \bibnamefont {Pfeiffer}}, \bibinfo {author} {\bibfnamefont {K.~W.}\ \bibnamefont {West}}, \bibinfo {author} {\bibfnamefont {K.~W.}\ \bibnamefont {Baldwin}},\ and\ \bibinfo {author} {\bibfnamefont {M.}~\bibnamefont {Shayegan}},\ }\bibfield  {title} {\bibinfo {title} {{Microwave spectroscopic observation of a Wigner solid within the $\ensuremath{\nu}=1/2$ fractional quantum Hall effect}},\ }\href {https://doi.org/10.1103/PhysRevB.95.045417} {\bibfield  {journal} {\bibinfo  {journal} {Phys. Rev. B}\ }\textbf {\bibinfo {volume} {95}},\ \bibinfo {pages} {045417} (\bibinfo {year} {2017})}\BibitemShut {NoStop}%
\bibitem [{\citenamefont {Peterson}\ \emph {et~al.}(2010)\citenamefont {Peterson}, \citenamefont {Papi\ifmmode~\acute{c}\else \'{c}\fi{}},\ and\ \citenamefont {Das~Sarma}}]{Peterson2.PRB.2010}%
  \BibitemOpen
  \bibfield  {author} {\bibinfo {author} {\bibfnamefont {M.~R.}\ \bibnamefont {Peterson}}, \bibinfo {author} {\bibfnamefont {Z.}~\bibnamefont {Papi\ifmmode~\acute{c}\else \'{c}\fi{}}},\ and\ \bibinfo {author} {\bibfnamefont {S.}~\bibnamefont {Das~Sarma}},\ }\bibfield  {title} {\bibinfo {title} {{Fractional quantum Hall effects in bilayers in the presence of interlayer tunneling and charge imbalance}},\ }\href {https://doi.org/10.1103/PhysRevB.82.235312} {\bibfield  {journal} {\bibinfo  {journal} {Phys. Rev. B}\ }\textbf {\bibinfo {volume} {82}},\ \bibinfo {pages} {235312} (\bibinfo {year} {2010})}\BibitemShut {NoStop}%
\bibitem [{\citenamefont {Willett}\ \emph {et~al.}(1988)\citenamefont {Willett}, \citenamefont {Stormer}, \citenamefont {Tsui}, \citenamefont {Gossard},\ and\ \citenamefont {English}}]{Willett.PRB.1988}%
  \BibitemOpen
  \bibfield  {author} {\bibinfo {author} {\bibfnamefont {R.~L.}\ \bibnamefont {Willett}}, \bibinfo {author} {\bibfnamefont {H.~L.}\ \bibnamefont {Stormer}}, \bibinfo {author} {\bibfnamefont {D.~C.}\ \bibnamefont {Tsui}}, \bibinfo {author} {\bibfnamefont {A.~C.}\ \bibnamefont {Gossard}},\ and\ \bibinfo {author} {\bibfnamefont {J.~H.}\ \bibnamefont {English}},\ }\bibfield  {title} {\bibinfo {title} {{Quantitative experimental test for the theoretical gap energies in the fractional quantum Hall effect}},\ }\href {https://doi.org/10.1103/PhysRevB.37.8476} {\bibfield  {journal} {\bibinfo  {journal} {Phys. Rev. B}\ }\textbf {\bibinfo {volume} {37}},\ \bibinfo {pages} {8476} (\bibinfo {year} {1988})}\BibitemShut {NoStop}%
\bibitem [{\citenamefont {Melik-Alaverdian}\ and\ \citenamefont {Bonesteel}(1995)}]{Melik-Alaverdian.PRB.1995}%
  \BibitemOpen
  \bibfield  {author} {\bibinfo {author} {\bibfnamefont {V.}~\bibnamefont {Melik-Alaverdian}}\ and\ \bibinfo {author} {\bibfnamefont {N.~E.}\ \bibnamefont {Bonesteel}},\ }\bibfield  {title} {\bibinfo {title} {{Composite fermions and Landau-level mixing in the fractional quantum Hall effect}},\ }\href {https://doi.org/10.1103/PhysRevB.52.R17032} {\bibfield  {journal} {\bibinfo  {journal} {Phys. Rev. B}\ }\textbf {\bibinfo {volume} {52}},\ \bibinfo {pages} {R17032} (\bibinfo {year} {1995})}\BibitemShut {NoStop}%
\bibitem [{\citenamefont {Park}\ \emph {et~al.}(1999)\citenamefont {Park}, \citenamefont {Meskini},\ and\ \citenamefont {Jain}}]{Park.JPhys.1999}%
  \BibitemOpen
  \bibfield  {author} {\bibinfo {author} {\bibfnamefont {K.}~\bibnamefont {Park}}, \bibinfo {author} {\bibfnamefont {N.}~\bibnamefont {Meskini}},\ and\ \bibinfo {author} {\bibfnamefont {J.}~\bibnamefont {Jain}},\ }\bibfield  {title} {\bibinfo {title} {{Activation gaps for the fractional quantum Hall effect: realistic treatment of transverse thickness}},\ }\href {https://doi.org/10.1088/0953-8984/11/38/308} {\bibfield  {journal} {\bibinfo  {journal} {J. Phys.: Condens. Matter}\ }\textbf {\bibinfo {volume} {11}},\ \bibinfo {pages} {7283} (\bibinfo {year} {1999})}\BibitemShut {NoStop}%
\bibitem [{\citenamefont {Morf}\ \emph {et~al.}(2002)\citenamefont {Morf}, \citenamefont {d'Ambrumenil},\ and\ \citenamefont {Das~Sarma}}]{Morf.PRB.2002}%
  \BibitemOpen
  \bibfield  {author} {\bibinfo {author} {\bibfnamefont {R.~H.}\ \bibnamefont {Morf}}, \bibinfo {author} {\bibfnamefont {N.}~\bibnamefont {d'Ambrumenil}},\ and\ \bibinfo {author} {\bibfnamefont {S.}~\bibnamefont {Das~Sarma}},\ }\bibfield  {title} {\bibinfo {title} {{Excitation gaps in fractional quantum Hall states: An exact diagonalization study}},\ }\href {https://doi.org/10.1103/PhysRevB.66.075408} {\bibfield  {journal} {\bibinfo  {journal} {Phys. Rev. B}\ }\textbf {\bibinfo {volume} {66}},\ \bibinfo {pages} {075408} (\bibinfo {year} {2002})}\BibitemShut {NoStop}%
\bibitem [{\citenamefont {Storni}\ \emph {et~al.}(2010)\citenamefont {Storni}, \citenamefont {Morf},\ and\ \citenamefont {Das~Sarma}}]{Storni.PRL.2010}%
  \BibitemOpen
  \bibfield  {author} {\bibinfo {author} {\bibfnamefont {M.}~\bibnamefont {Storni}}, \bibinfo {author} {\bibfnamefont {R.~H.}\ \bibnamefont {Morf}},\ and\ \bibinfo {author} {\bibfnamefont {S.}~\bibnamefont {Das~Sarma}},\ }\bibfield  {title} {\bibinfo {title} {{Fractional Quantum Hall State at $\ensuremath{\nu}=\frac{5}{2}$ and the Moore-Read Pfaffian}},\ }\href {https://doi.org/10.1103/PhysRevLett.104.076803} {\bibfield  {journal} {\bibinfo  {journal} {Phys. Rev. Lett.}\ }\textbf {\bibinfo {volume} {104}},\ \bibinfo {pages} {076803} (\bibinfo {year} {2010})}\BibitemShut {NoStop}%
\bibitem [{\citenamefont {Villegas~Rosales}\ \emph {et~al.}(2021)\citenamefont {Villegas~Rosales}, \citenamefont {Madathil}, \citenamefont {Chung}, \citenamefont {Pfeiffer}, \citenamefont {West}, \citenamefont {Baldwin},\ and\ \citenamefont {Shayegan}}]{VillegasRosales.PRL.2021}%
  \BibitemOpen
  \bibfield  {author} {\bibinfo {author} {\bibfnamefont {K.~A.}\ \bibnamefont {Villegas~Rosales}}, \bibinfo {author} {\bibfnamefont {P.~T.}\ \bibnamefont {Madathil}}, \bibinfo {author} {\bibfnamefont {Y.~J.}\ \bibnamefont {Chung}}, \bibinfo {author} {\bibfnamefont {L.~N.}\ \bibnamefont {Pfeiffer}}, \bibinfo {author} {\bibfnamefont {K.~W.}\ \bibnamefont {West}}, \bibinfo {author} {\bibfnamefont {K.~W.}\ \bibnamefont {Baldwin}},\ and\ \bibinfo {author} {\bibfnamefont {M.}~\bibnamefont {Shayegan}},\ }\bibfield  {title} {\bibinfo {title} {{Fractional Quantum Hall Effect Energy Gaps: Role of Electron Layer Thickness}},\ }\href {https://doi.org/10.1103/PhysRevLett.127.056801} {\bibfield  {journal} {\bibinfo  {journal} {Phys. Rev. Lett.}\ }\textbf {\bibinfo {volume} {127}},\ \bibinfo {pages} {056801} (\bibinfo {year} {2021})}\BibitemShut {NoStop}%
\bibitem [{\citenamefont {Ho}(1995)}]{Ho.PRL.1995}%
  \BibitemOpen
  \bibfield  {author} {\bibinfo {author} {\bibfnamefont {T.-L.}\ \bibnamefont {Ho}},\ }\bibfield  {title} {\bibinfo {title} {{Broken Symmetry of Two-Component $\mathit{\ensuremath{\nu}}=1/2$ Quantum Hall States}},\ }\href {https://doi.org/10.1103/PhysRevLett.75.1186} {\bibfield  {journal} {\bibinfo  {journal} {Phys. Rev. Lett.}\ }\textbf {\bibinfo {volume} {75}},\ \bibinfo {pages} {1186} (\bibinfo {year} {1995})}\BibitemShut {NoStop}%
\bibitem [{\citenamefont {Halperin}(1994)}]{Halperin.Surf.Sci.1994}%
  \BibitemOpen
  \bibfield  {author} {\bibinfo {author} {\bibfnamefont {B.~I.}\ \bibnamefont {Halperin}},\ }\bibfield  {title} {\bibinfo {title} {{Theories for $\nu = 1/2$ in single- and double-layer systems}},\ }\href {https://doi.org/https://doi.org/10.1016/0039-6028(94)90850-8} {\bibfield  {journal} {\bibinfo  {journal} {Surf. Sci.}\ }\textbf {\bibinfo {volume} {305}},\ \bibinfo {pages} {1} (\bibinfo {year} {1994})}\BibitemShut {NoStop}%
\bibitem [{\citenamefont {Read}\ and\ \citenamefont {Green}(2000)}]{Read.PRB.2000}%
  \BibitemOpen
  \bibfield  {author} {\bibinfo {author} {\bibfnamefont {N.}~\bibnamefont {Read}}\ and\ \bibinfo {author} {\bibfnamefont {D.}~\bibnamefont {Green}},\ }\bibfield  {title} {\bibinfo {title} {{Paired states of fermions in two dimensions with breaking of parity and time-reversal symmetries and the fractional quantum Hall effect}},\ }\href {https://doi.org/10.1103/PhysRevB.61.10267} {\bibfield  {journal} {\bibinfo  {journal} {Phys. Rev. B}\ }\textbf {\bibinfo {volume} {61}},\ \bibinfo {pages} {10267} (\bibinfo {year} {2000})}\BibitemShut {NoStop}%
\bibitem [{\citenamefont {Wen}(2000)}]{Wen.PRL.2000}%
  \BibitemOpen
  \bibfield  {author} {\bibinfo {author} {\bibfnamefont {X.-G.}\ \bibnamefont {Wen}},\ }\bibfield  {title} {\bibinfo {title} {{Continuous Topological Phase Transitions between Clean Quantum Hall States}},\ }\href {https://doi.org/10.1103/PhysRevLett.84.3950} {\bibfield  {journal} {\bibinfo  {journal} {Phys. Rev. Lett.}\ }\textbf {\bibinfo {volume} {84}},\ \bibinfo {pages} {3950} (\bibinfo {year} {2000})}\BibitemShut {NoStop}%
\bibitem [{\citenamefont {Yang}(2017)}]{Yang.PRB.2017}%
  \BibitemOpen
  \bibfield  {author} {\bibinfo {author} {\bibfnamefont {K.}~\bibnamefont {Yang}},\ }\bibfield  {title} {\bibinfo {title} {{Interface and phase transition between Moore-Read and Halperin 331 fractional quantum Hall states: Realization of chiral Majorana fermion}},\ }\href {https://doi.org/10.1103/PhysRevB.96.241305} {\bibfield  {journal} {\bibinfo  {journal} {Phys. Rev. B}\ }\textbf {\bibinfo {volume} {96}},\ \bibinfo {pages} {241305} (\bibinfo {year} {2017})}\BibitemShut {NoStop}%
\bibitem [{\citenamefont {Ma}\ and\ \citenamefont {Yang}(2022)}]{Ma.PRB.2022}%
  \BibitemOpen
  \bibfield  {author} {\bibinfo {author} {\bibfnamefont {K.~K.~W.}\ \bibnamefont {Ma}}\ and\ \bibinfo {author} {\bibfnamefont {K.}~\bibnamefont {Yang}},\ }\bibfield  {title} {\bibinfo {title} {{Simple analog of the black-hole information paradox in quantum Hall interfaces}},\ }\href {https://doi.org/10.1103/PhysRevB.105.045306} {\bibfield  {journal} {\bibinfo  {journal} {Phys. Rev. B}\ }\textbf {\bibinfo {volume} {105}},\ \bibinfo {pages} {045306} (\bibinfo {year} {2022})}\BibitemShut {NoStop}%
\bibitem [{\citenamefont {Pan}\ \emph {et~al.}(1999)\citenamefont {Pan}, \citenamefont {Xia}, \citenamefont {Shvarts}, \citenamefont {Adams}, \citenamefont {Stormer}, \citenamefont {Tsui}, \citenamefont {Pfeiffer}, \citenamefont {Baldwin},\ and\ \citenamefont {West}}]{Pan.PRL.1999}%
  \BibitemOpen
  \bibfield  {author} {\bibinfo {author} {\bibfnamefont {W.}~\bibnamefont {Pan}}, \bibinfo {author} {\bibfnamefont {J.-S.}\ \bibnamefont {Xia}}, \bibinfo {author} {\bibfnamefont {V.}~\bibnamefont {Shvarts}}, \bibinfo {author} {\bibfnamefont {D.~E.}\ \bibnamefont {Adams}}, \bibinfo {author} {\bibfnamefont {H.~L.}\ \bibnamefont {Stormer}}, \bibinfo {author} {\bibfnamefont {D.~C.}\ \bibnamefont {Tsui}}, \bibinfo {author} {\bibfnamefont {L.~N.}\ \bibnamefont {Pfeiffer}}, \bibinfo {author} {\bibfnamefont {K.~W.}\ \bibnamefont {Baldwin}},\ and\ \bibinfo {author} {\bibfnamefont {K.~W.}\ \bibnamefont {West}},\ }\bibfield  {title} {\bibinfo {title} {{Exact Quantization of the Even-Denominator Fractional Quantum Hall State at $\mathit{\ensuremath{\nu}}\phantom{\rule{0ex}{0ex}}=\phantom{\rule{0ex}{0ex}}5/2$ Landau Level Filling Factor}},\ }\href {https://doi.org/10.1103/PhysRevLett.83.3530} {\bibfield  {journal} {\bibinfo  {journal} {Phys. Rev. Lett.}\ }\textbf {\bibinfo {volume} {83}},\ \bibinfo {pages} {3530}
  (\bibinfo {year} {1999})}\BibitemShut {NoStop}%
\bibitem [{\citenamefont {Chung}\ \emph {et~al.}(2021)\citenamefont {Chung}, \citenamefont {Villegas~Rosales}, \citenamefont {Baldwin}, \citenamefont {Madathil}, \citenamefont {West}, \citenamefont {Shayegan},\ and\ \citenamefont {Pfeiffer}}]{Chung.NatMater.2021}%
  \BibitemOpen
  \bibfield  {author} {\bibinfo {author} {\bibfnamefont {Y.~J.}\ \bibnamefont {Chung}}, \bibinfo {author} {\bibfnamefont {K.}~\bibnamefont {Villegas~Rosales}}, \bibinfo {author} {\bibfnamefont {K.}~\bibnamefont {Baldwin}}, \bibinfo {author} {\bibfnamefont {P.}~\bibnamefont {Madathil}}, \bibinfo {author} {\bibfnamefont {K.}~\bibnamefont {West}}, \bibinfo {author} {\bibfnamefont {M.}~\bibnamefont {Shayegan}},\ and\ \bibinfo {author} {\bibfnamefont {L.}~\bibnamefont {Pfeiffer}},\ }\bibfield  {title} {\bibinfo {title} {{Ultra-high-quality two-dimensional electron systems}},\ }\href {https://doi.org/10.1038/s41563-021-00942-3} {\bibfield  {journal} {\bibinfo  {journal} {Nat. Mater.}\ }\textbf {\bibinfo {volume} {20}},\ \bibinfo {pages} {632} (\bibinfo {year} {2021})}\BibitemShut {NoStop}%
\bibitem [{\citenamefont {Falson}\ \emph {et~al.}(2015)\citenamefont {Falson}, \citenamefont {Maryenko}, \citenamefont {Friess}, \citenamefont {Zhang}, \citenamefont {Kozuka}, \citenamefont {Tsukazaki}, \citenamefont {Smet},\ and\ \citenamefont {Kawasaki}}]{Falson.Nat.Phys.2015}%
  \BibitemOpen
  \bibfield  {author} {\bibinfo {author} {\bibfnamefont {J.}~\bibnamefont {Falson}}, \bibinfo {author} {\bibfnamefont {D.}~\bibnamefont {Maryenko}}, \bibinfo {author} {\bibfnamefont {B.}~\bibnamefont {Friess}}, \bibinfo {author} {\bibfnamefont {D.}~\bibnamefont {Zhang}}, \bibinfo {author} {\bibfnamefont {Y.}~\bibnamefont {Kozuka}}, \bibinfo {author} {\bibfnamefont {A.}~\bibnamefont {Tsukazaki}}, \bibinfo {author} {\bibfnamefont {J.}~\bibnamefont {Smet}},\ and\ \bibinfo {author} {\bibfnamefont {M.}~\bibnamefont {Kawasaki}},\ }\bibfield  {title} {\bibinfo {title} {{Even-denominator fractional quantum Hall physics in ZnO}},\ }\href {https://doi.org/https://doi.org/10.1038/nphys3259} {\bibfield  {journal} {\bibinfo  {journal} {Nat. Phys.}\ }\textbf {\bibinfo {volume} {11}},\ \bibinfo {pages} {347} (\bibinfo {year} {2015})}\BibitemShut {NoStop}%
\bibitem [{\citenamefont {Hossain}\ \emph {et~al.}(2018)\citenamefont {Hossain}, \citenamefont {Ma}, \citenamefont {Chung}, \citenamefont {Pfeiffer}, \citenamefont {West}, \citenamefont {Baldwin},\ and\ \citenamefont {Shayegan}}]{Shafayat.PRL.2018}%
  \BibitemOpen
  \bibfield  {author} {\bibinfo {author} {\bibfnamefont {M.~S.}\ \bibnamefont {Hossain}}, \bibinfo {author} {\bibfnamefont {M.~K.}\ \bibnamefont {Ma}}, \bibinfo {author} {\bibfnamefont {Y.~J.}\ \bibnamefont {Chung}}, \bibinfo {author} {\bibfnamefont {L.~N.}\ \bibnamefont {Pfeiffer}}, \bibinfo {author} {\bibfnamefont {K.~W.}\ \bibnamefont {West}}, \bibinfo {author} {\bibfnamefont {K.~W.}\ \bibnamefont {Baldwin}},\ and\ \bibinfo {author} {\bibfnamefont {M.}~\bibnamefont {Shayegan}},\ }\bibfield  {title} {\bibinfo {title} {{Unconventional Anisotropic Even-Denominator Fractional Quantum Hall State in a System with Mass Anisotropy}},\ }\href {https://doi.org/10.1103/PhysRevLett.121.256601} {\bibfield  {journal} {\bibinfo  {journal} {Phys. Rev. Lett.}\ }\textbf {\bibinfo {volume} {121}},\ \bibinfo {pages} {256601} (\bibinfo {year} {2018})}\BibitemShut {NoStop}%
\bibitem [{\citenamefont {Ki}\ \emph {et~al.}(2014)\citenamefont {Ki}, \citenamefont {Fal’ko}, \citenamefont {Abanin},\ and\ \citenamefont {Morpurgo}}]{Ki.NanoLett.2014}%
  \BibitemOpen
  \bibfield  {author} {\bibinfo {author} {\bibfnamefont {D.-K.}\ \bibnamefont {Ki}}, \bibinfo {author} {\bibfnamefont {V.~I.}\ \bibnamefont {Fal’ko}}, \bibinfo {author} {\bibfnamefont {D.~A.}\ \bibnamefont {Abanin}},\ and\ \bibinfo {author} {\bibfnamefont {A.~F.}\ \bibnamefont {Morpurgo}},\ }\bibfield  {title} {\bibinfo {title} {{Observation of even denominator fractional quantum Hall effect in suspended bilayer graphene}},\ }\href {https://doi.org/10.1021/nl5003922} {\bibfield  {journal} {\bibinfo  {journal} {Nano Lett.}\ }\textbf {\bibinfo {volume} {14}},\ \bibinfo {pages} {2135} (\bibinfo {year} {2014})}\BibitemShut {NoStop}%
\bibitem [{\citenamefont {Dean}\ \emph {et~al.}()\citenamefont {Dean}, \citenamefont {Kim}, \citenamefont {Li},\ and\ \citenamefont {Young}}]{Review.Dean.Kim.Li.Young.2020}%
  \BibitemOpen
  \bibfield  {author} {\bibinfo {author} {\bibfnamefont {C.}~\bibnamefont {Dean}}, \bibinfo {author} {\bibfnamefont {P.}~\bibnamefont {Kim}}, \bibinfo {author} {\bibfnamefont {J.~I.~A.}\ \bibnamefont {Li}},\ and\ \bibinfo {author} {\bibfnamefont {A.}~\bibnamefont {Young}},\ }\bibinfo {title} {{Fractional Quantum Hall Effects in Graphene}},\ in\ \href {https://doi.org/10.1142/9789811217494_0007} {\emph {\bibinfo {booktitle} {Fractional Quantum Hall Effects}}},\ \bibinfo {editor} {edited by\ \bibinfo {editor} {\bibfnamefont {B.~I.}\ \bibnamefont {Halperin}}\ and\ \bibinfo {editor} {\bibfnamefont {J.~K.}\ \bibnamefont {Jain}}},\ Chap.~\bibinfo {chapter} {7}, pp.\ \bibinfo {pages} {317--375}\BibitemShut {NoStop}%
\bibitem [{\citenamefont {Kim}\ \emph {et~al.}(2019)\citenamefont {Kim}, \citenamefont {Balram}, \citenamefont {Taniguchi}, \citenamefont {Watanabe}, \citenamefont {Jain},\ and\ \citenamefont {Smet}}]{Kim.NatPhys.2019}%
  \BibitemOpen
  \bibfield  {author} {\bibinfo {author} {\bibfnamefont {Y.}~\bibnamefont {Kim}}, \bibinfo {author} {\bibfnamefont {A.~C.}\ \bibnamefont {Balram}}, \bibinfo {author} {\bibfnamefont {T.}~\bibnamefont {Taniguchi}}, \bibinfo {author} {\bibfnamefont {K.}~\bibnamefont {Watanabe}}, \bibinfo {author} {\bibfnamefont {J.~K.}\ \bibnamefont {Jain}},\ and\ \bibinfo {author} {\bibfnamefont {J.~H.}\ \bibnamefont {Smet}},\ }\bibfield  {title} {\bibinfo {title} {{Even denominator fractional quantum Hall states in higher Landau levels of graphene}},\ }\href {https://doi.org/10.1038/s41567-018-0355-x} {\bibfield  {journal} {\bibinfo  {journal} {Nat. Phys.}\ }\textbf {\bibinfo {volume} {15}},\ \bibinfo {pages} {154} (\bibinfo {year} {2019})}\BibitemShut {NoStop}%
\bibitem [{\citenamefont {Huang}\ \emph {et~al.}(2022)\citenamefont {Huang}, \citenamefont {Fu}, \citenamefont {Hickey}, \citenamefont {Alem}, \citenamefont {Lin}, \citenamefont {Watanabe}, \citenamefont {Taniguchi},\ and\ \citenamefont {Zhu}}]{Huang.PRX.2022}%
  \BibitemOpen
  \bibfield  {author} {\bibinfo {author} {\bibfnamefont {K.}~\bibnamefont {Huang}}, \bibinfo {author} {\bibfnamefont {H.}~\bibnamefont {Fu}}, \bibinfo {author} {\bibfnamefont {D.~R.}\ \bibnamefont {Hickey}}, \bibinfo {author} {\bibfnamefont {N.}~\bibnamefont {Alem}}, \bibinfo {author} {\bibfnamefont {X.}~\bibnamefont {Lin}}, \bibinfo {author} {\bibfnamefont {K.}~\bibnamefont {Watanabe}}, \bibinfo {author} {\bibfnamefont {T.}~\bibnamefont {Taniguchi}},\ and\ \bibinfo {author} {\bibfnamefont {J.}~\bibnamefont {Zhu}},\ }\bibfield  {title} {\bibinfo {title} {{Valley Isospin Controlled Fractional Quantum Hall States in Bilayer Graphene}},\ }\href {https://doi.org/10.1103/PhysRevX.12.031019} {\bibfield  {journal} {\bibinfo  {journal} {Phys. Rev. X}\ }\textbf {\bibinfo {volume} {12}},\ \bibinfo {pages} {031019} (\bibinfo {year} {2022})}\BibitemShut {NoStop}%
\bibitem [{\citenamefont {Assouline}\ \emph {et~al.}(2024)\citenamefont {Assouline}, \citenamefont {Wang}, \citenamefont {Zhou}, \citenamefont {Cohen}, \citenamefont {Yang}, \citenamefont {Zhang}, \citenamefont {Taniguchi}, \citenamefont {Watanabe}, \citenamefont {Mong}, \citenamefont {Zaletel},\ and\ \citenamefont {Young}}]{Assouline.PRL.2024}%
  \BibitemOpen
  \bibfield  {author} {\bibinfo {author} {\bibfnamefont {A.}~\bibnamefont {Assouline}}, \bibinfo {author} {\bibfnamefont {T.}~\bibnamefont {Wang}}, \bibinfo {author} {\bibfnamefont {H.}~\bibnamefont {Zhou}}, \bibinfo {author} {\bibfnamefont {L.~A.}\ \bibnamefont {Cohen}}, \bibinfo {author} {\bibfnamefont {F.}~\bibnamefont {Yang}}, \bibinfo {author} {\bibfnamefont {R.}~\bibnamefont {Zhang}}, \bibinfo {author} {\bibfnamefont {T.}~\bibnamefont {Taniguchi}}, \bibinfo {author} {\bibfnamefont {K.}~\bibnamefont {Watanabe}}, \bibinfo {author} {\bibfnamefont {R.~S.~K.}\ \bibnamefont {Mong}}, \bibinfo {author} {\bibfnamefont {M.~P.}\ \bibnamefont {Zaletel}},\ and\ \bibinfo {author} {\bibfnamefont {A.~F.}\ \bibnamefont {Young}},\ }\bibfield  {title} {\bibinfo {title} {{Energy Gap of the Even-Denominator Fractional Quantum Hall State in Bilayer Graphene}},\ }\href {https://doi.org/10.1103/PhysRevLett.132.046603} {\bibfield  {journal} {\bibinfo  {journal} {Phys. Rev. Lett.}\ }\textbf {\bibinfo {volume} {132}},\ \bibinfo
  {pages} {046603} (\bibinfo {year} {2024})}\BibitemShut {NoStop}%
\bibitem [{\citenamefont {Kumar}\ \emph {et~al.}(2025)\citenamefont {Kumar}, \citenamefont {Haug}, \citenamefont {Kim}, \citenamefont {Yutushui}, \citenamefont {Khudiakov}, \citenamefont {Bhardwaj}, \citenamefont {Ilin}, \citenamefont {Watanabe}, \citenamefont {Taniguchi}, \citenamefont {Mross},\ and\ \citenamefont {Ronen}}]{Kumar.Nat.Comm.2025}%
  \BibitemOpen
  \bibfield  {author} {\bibinfo {author} {\bibfnamefont {R.}~\bibnamefont {Kumar}}, \bibinfo {author} {\bibfnamefont {A.}~\bibnamefont {Haug}}, \bibinfo {author} {\bibfnamefont {J.}~\bibnamefont {Kim}}, \bibinfo {author} {\bibfnamefont {M.}~\bibnamefont {Yutushui}}, \bibinfo {author} {\bibfnamefont {K.}~\bibnamefont {Khudiakov}}, \bibinfo {author} {\bibfnamefont {V.}~\bibnamefont {Bhardwaj}}, \bibinfo {author} {\bibfnamefont {A.}~\bibnamefont {Ilin}}, \bibinfo {author} {\bibfnamefont {K.}~\bibnamefont {Watanabe}}, \bibinfo {author} {\bibfnamefont {T.}~\bibnamefont {Taniguchi}}, \bibinfo {author} {\bibfnamefont {D.~F.}\ \bibnamefont {Mross}},\ and\ \bibinfo {author} {\bibfnamefont {Y.}~\bibnamefont {Ronen}},\ }\bibfield  {title} {\bibinfo {title} {{Quarter- and half-filled quantum Hall states and their competing interactions in bilayer graphene}},\ }\href {https://doi.org/10.1038/s41467-025-62650-9} {\bibfield  {journal} {\bibinfo  {journal} {Nat. Comm.}\ }\textbf {\bibinfo {volume} {16}},\ \bibinfo {pages}
  {7255} (\bibinfo {year} {2025})}\BibitemShut {NoStop}%
\bibitem [{\citenamefont {Hu}\ \emph {et~al.}(2025)\citenamefont {Hu}, \citenamefont {Tsui}, \citenamefont {He}, \citenamefont {Kamber}, \citenamefont {Wang}, \citenamefont {Mohammadi}, \citenamefont {Watanabe}, \citenamefont {Taniguchi}, \citenamefont {Papi{\'c}}, \citenamefont {Zaletel} \emph {et~al.}}]{Hu.Nat.Phys.2025}%
  \BibitemOpen
  \bibfield  {author} {\bibinfo {author} {\bibfnamefont {Y.}~\bibnamefont {Hu}}, \bibinfo {author} {\bibfnamefont {Y.-C.}\ \bibnamefont {Tsui}}, \bibinfo {author} {\bibfnamefont {M.}~\bibnamefont {He}}, \bibinfo {author} {\bibfnamefont {U.}~\bibnamefont {Kamber}}, \bibinfo {author} {\bibfnamefont {T.}~\bibnamefont {Wang}}, \bibinfo {author} {\bibfnamefont {A.~S.}\ \bibnamefont {Mohammadi}}, \bibinfo {author} {\bibfnamefont {K.}~\bibnamefont {Watanabe}}, \bibinfo {author} {\bibfnamefont {T.}~\bibnamefont {Taniguchi}}, \bibinfo {author} {\bibfnamefont {Z.}~\bibnamefont {Papi{\'c}}}, \bibinfo {author} {\bibfnamefont {M.~P.}\ \bibnamefont {Zaletel}}, \emph {et~al.},\ }\bibfield  {title} {\bibinfo {title} {{High-resolution tunnelling spectroscopy of fractional quantum Hall states}},\ }\href {https://doi.org/10.1038/s41567-025-02830-y} {\bibfield  {journal} {\bibinfo  {journal} {Nat. Phys.}\ }\textbf {\bibinfo {volume} {21}},\ \bibinfo {pages} {716–} (\bibinfo {year} {2025})}\BibitemShut {NoStop}%
\bibitem [{\citenamefont {Chanda}\ \emph {et~al.}(2025)\citenamefont {Chanda}, \citenamefont {Kaur}, \citenamefont {Singh}, \citenamefont {Watanabe}, \citenamefont {Taniguchi}, \citenamefont {Jain}, \citenamefont {Khanna}, \citenamefont {Balram},\ and\ \citenamefont {Bid}}]{Chanda.Preprint.2024}%
  \BibitemOpen
  \bibfield  {author} {\bibinfo {author} {\bibfnamefont {T.}~\bibnamefont {Chanda}}, \bibinfo {author} {\bibfnamefont {S.}~\bibnamefont {Kaur}}, \bibinfo {author} {\bibfnamefont {H.}~\bibnamefont {Singh}}, \bibinfo {author} {\bibfnamefont {K.}~\bibnamefont {Watanabe}}, \bibinfo {author} {\bibfnamefont {T.}~\bibnamefont {Taniguchi}}, \bibinfo {author} {\bibfnamefont {M.}~\bibnamefont {Jain}}, \bibinfo {author} {\bibfnamefont {U.}~\bibnamefont {Khanna}}, \bibinfo {author} {\bibfnamefont {A.~C.}\ \bibnamefont {Balram}},\ and\ \bibinfo {author} {\bibfnamefont {A.}~\bibnamefont {Bid}},\ }\bibfield  {title} {\bibinfo {title} {{Even denominator fractional quantum Hall states in the zeroth Landau level of monolayer-like band of ABA trilayer graphene}},\ }\Eprint {https://arxiv.org/abs/2502.06245} {arXiv:2502.06245 [cond-mat.mes-hall]}  (\bibinfo {year} {2025})\BibitemShut {NoStop}%
\bibitem [{\citenamefont {Shi}\ \emph {et~al.}(2020)\citenamefont {Shi}, \citenamefont {Shih}, \citenamefont {Gustafsson}, \citenamefont {Rhodes}, \citenamefont {Kim}, \citenamefont {Watanabe}, \citenamefont {Taniguchi}, \citenamefont {Papi{\'c}}, \citenamefont {Hone},\ and\ \citenamefont {Dean}}]{Shi.NatureNanotech.2020}%
  \BibitemOpen
  \bibfield  {author} {\bibinfo {author} {\bibfnamefont {Q.}~\bibnamefont {Shi}}, \bibinfo {author} {\bibfnamefont {E.-M.}\ \bibnamefont {Shih}}, \bibinfo {author} {\bibfnamefont {M.~V.}\ \bibnamefont {Gustafsson}}, \bibinfo {author} {\bibfnamefont {D.~A.}\ \bibnamefont {Rhodes}}, \bibinfo {author} {\bibfnamefont {B.}~\bibnamefont {Kim}}, \bibinfo {author} {\bibfnamefont {K.}~\bibnamefont {Watanabe}}, \bibinfo {author} {\bibfnamefont {T.}~\bibnamefont {Taniguchi}}, \bibinfo {author} {\bibfnamefont {Z.}~\bibnamefont {Papi{\'c}}}, \bibinfo {author} {\bibfnamefont {J.}~\bibnamefont {Hone}},\ and\ \bibinfo {author} {\bibfnamefont {C.~R.}\ \bibnamefont {Dean}},\ }\bibfield  {title} {\bibinfo {title} {{Odd-and even-denominator fractional quantum Hall states in monolayer WSe\textsubscript{2}}},\ }\href {https://doi.org/10.1038/s41565-020-0685-6} {\bibfield  {journal} {\bibinfo  {journal} {Nat. Nanotechnol.}\ }\textbf {\bibinfo {volume} {15}},\ \bibinfo {pages} {569} (\bibinfo {year} {2020})}\BibitemShut {NoStop}%
\bibitem [{\citenamefont {Liu}\ \emph {et~al.}(2014{\natexlab{a}})\citenamefont {Liu}, \citenamefont {Graninger}, \citenamefont {Hasdemir}, \citenamefont {Shayegan}, \citenamefont {Pfeiffer}, \citenamefont {West}, \citenamefont {Baldwin},\ and\ \citenamefont {Winkler}}]{Liu.PRL.2014}%
  \BibitemOpen
  \bibfield  {author} {\bibinfo {author} {\bibfnamefont {Y.}~\bibnamefont {Liu}}, \bibinfo {author} {\bibfnamefont {A.~L.}\ \bibnamefont {Graninger}}, \bibinfo {author} {\bibfnamefont {S.}~\bibnamefont {Hasdemir}}, \bibinfo {author} {\bibfnamefont {M.}~\bibnamefont {Shayegan}}, \bibinfo {author} {\bibfnamefont {L.~N.}\ \bibnamefont {Pfeiffer}}, \bibinfo {author} {\bibfnamefont {K.~W.}\ \bibnamefont {West}}, \bibinfo {author} {\bibfnamefont {K.~W.}\ \bibnamefont {Baldwin}},\ and\ \bibinfo {author} {\bibfnamefont {R.}~\bibnamefont {Winkler}},\ }\bibfield  {title} {\bibinfo {title} {{Fractional Quantum Hall Effect at $\ensuremath{\nu}=1/2$ in Hole Systems Confined to GaAs Quantum Wells}},\ }\href {https://doi.org/10.1103/PhysRevLett.112.046804} {\bibfield  {journal} {\bibinfo  {journal} {Phys. Rev. Lett.}\ }\textbf {\bibinfo {volume} {112}},\ \bibinfo {pages} {046804} (\bibinfo {year} {2014}{\natexlab{a}})}\BibitemShut {NoStop}%
\bibitem [{\citenamefont {Liu}\ \emph {et~al.}(2014{\natexlab{b}})\citenamefont {Liu}, \citenamefont {Hasdemir}, \citenamefont {Kamburov}, \citenamefont {Graninger}, \citenamefont {Shayegan}, \citenamefont {Pfeiffer}, \citenamefont {West}, \citenamefont {Baldwin},\ and\ \citenamefont {Winkler}}]{Liu.PRB.2014}%
  \BibitemOpen
  \bibfield  {author} {\bibinfo {author} {\bibfnamefont {Y.}~\bibnamefont {Liu}}, \bibinfo {author} {\bibfnamefont {S.}~\bibnamefont {Hasdemir}}, \bibinfo {author} {\bibfnamefont {D.}~\bibnamefont {Kamburov}}, \bibinfo {author} {\bibfnamefont {A.~L.}\ \bibnamefont {Graninger}}, \bibinfo {author} {\bibfnamefont {M.}~\bibnamefont {Shayegan}}, \bibinfo {author} {\bibfnamefont {L.~N.}\ \bibnamefont {Pfeiffer}}, \bibinfo {author} {\bibfnamefont {K.~W.}\ \bibnamefont {West}}, \bibinfo {author} {\bibfnamefont {K.~W.}\ \bibnamefont {Baldwin}},\ and\ \bibinfo {author} {\bibfnamefont {R.}~\bibnamefont {Winkler}},\ }\bibfield  {title} {\bibinfo {title} {{Even-denominator fractional quantum Hall effect at a Landau level crossing}},\ }\href {https://doi.org/10.1103/PhysRevB.89.165313} {\bibfield  {journal} {\bibinfo  {journal} {Phys. Rev. B}\ }\textbf {\bibinfo {volume} {89}},\ \bibinfo {pages} {165313} (\bibinfo {year} {2014}{\natexlab{b}})}\BibitemShut {NoStop}%
\bibitem [{\citenamefont {Hossain}\ \emph {et~al.}(2023)\citenamefont {Hossain}, \citenamefont {Ma}, \citenamefont {Chung}, \citenamefont {Singh}, \citenamefont {Gupta}, \citenamefont {West}, \citenamefont {Baldwin}, \citenamefont {Pfeiffer}, \citenamefont {Winkler},\ and\ \citenamefont {Shayegan}}]{Shafayat.PRL.2023}%
  \BibitemOpen
  \bibfield  {author} {\bibinfo {author} {\bibfnamefont {M.~S.}\ \bibnamefont {Hossain}}, \bibinfo {author} {\bibfnamefont {M.~K.}\ \bibnamefont {Ma}}, \bibinfo {author} {\bibfnamefont {Y.~J.}\ \bibnamefont {Chung}}, \bibinfo {author} {\bibfnamefont {S.~K.}\ \bibnamefont {Singh}}, \bibinfo {author} {\bibfnamefont {A.}~\bibnamefont {Gupta}}, \bibinfo {author} {\bibfnamefont {K.~W.}\ \bibnamefont {West}}, \bibinfo {author} {\bibfnamefont {K.~W.}\ \bibnamefont {Baldwin}}, \bibinfo {author} {\bibfnamefont {L.~N.}\ \bibnamefont {Pfeiffer}}, \bibinfo {author} {\bibfnamefont {R.}~\bibnamefont {Winkler}},\ and\ \bibinfo {author} {\bibfnamefont {M.}~\bibnamefont {Shayegan}},\ }\bibfield  {title} {\bibinfo {title} {{Valley-Tunable Even-Denominator Fractional Quantum Hall State in the Lowest Landau Level of an Anisotropic System}},\ }\href {https://doi.org/10.1103/PhysRevLett.130.126301} {\bibfield  {journal} {\bibinfo  {journal} {Phys. Rev. Lett.}\ }\textbf {\bibinfo {volume} {130}},\ \bibinfo {pages} {126301}
  (\bibinfo {year} {2023})}\BibitemShut {NoStop}%
\bibitem [{\citenamefont {Wang}\ \emph {et~al.}(2022)\citenamefont {Wang}, \citenamefont {Gupta}, \citenamefont {Singh}, \citenamefont {Chung}, \citenamefont {Pfeiffer}, \citenamefont {West}, \citenamefont {Baldwin}, \citenamefont {Winkler},\ and\ \citenamefont {Shayegan}}]{Wang.PRL.2022}%
  \BibitemOpen
  \bibfield  {author} {\bibinfo {author} {\bibfnamefont {C.}~\bibnamefont {Wang}}, \bibinfo {author} {\bibfnamefont {A.}~\bibnamefont {Gupta}}, \bibinfo {author} {\bibfnamefont {S.~K.}\ \bibnamefont {Singh}}, \bibinfo {author} {\bibfnamefont {Y.~J.}\ \bibnamefont {Chung}}, \bibinfo {author} {\bibfnamefont {L.~N.}\ \bibnamefont {Pfeiffer}}, \bibinfo {author} {\bibfnamefont {K.~W.}\ \bibnamefont {West}}, \bibinfo {author} {\bibfnamefont {K.~W.}\ \bibnamefont {Baldwin}}, \bibinfo {author} {\bibfnamefont {R.}~\bibnamefont {Winkler}},\ and\ \bibinfo {author} {\bibfnamefont {M.}~\bibnamefont {Shayegan}},\ }\bibfield  {title} {\bibinfo {title} {{Even-Denominator Fractional Quantum Hall State at Filling Factor $\ensuremath{\nu}=3/4$}},\ }\href {https://doi.org/10.1103/PhysRevLett.129.156801} {\bibfield  {journal} {\bibinfo  {journal} {Phys. Rev. Lett.}\ }\textbf {\bibinfo {volume} {129}},\ \bibinfo {pages} {156801} (\bibinfo {year} {2022})}\BibitemShut {NoStop}%
\bibitem [{\citenamefont {Wang}\ \emph {et~al.}(2023{\natexlab{a}})\citenamefont {Wang}, \citenamefont {Gupta}, \citenamefont {Madathil}, \citenamefont {Singh}, \citenamefont {Chung}, \citenamefont {Pfeiffer}, \citenamefont {Baldwin},\ and\ \citenamefont {Shayegan}}]{Wang.PNAS.2023}%
  \BibitemOpen
  \bibfield  {author} {\bibinfo {author} {\bibfnamefont {C.}~\bibnamefont {Wang}}, \bibinfo {author} {\bibfnamefont {A.}~\bibnamefont {Gupta}}, \bibinfo {author} {\bibfnamefont {P.~T.}\ \bibnamefont {Madathil}}, \bibinfo {author} {\bibfnamefont {S.~K.}\ \bibnamefont {Singh}}, \bibinfo {author} {\bibfnamefont {Y.~J.}\ \bibnamefont {Chung}}, \bibinfo {author} {\bibfnamefont {L.~N.}\ \bibnamefont {Pfeiffer}}, \bibinfo {author} {\bibfnamefont {K.~W.}\ \bibnamefont {Baldwin}},\ and\ \bibinfo {author} {\bibfnamefont {M.}~\bibnamefont {Shayegan}},\ }\bibfield  {title} {\bibinfo {title} {{Next-generation even-denominator fractional quantum Hall states of interacting composite fermions}},\ }\href {https://doi.org/10.1073/pnas.2314212120} {\bibfield  {journal} {\bibinfo  {journal} {Proc. Natl. Acad. Sci. U.S.A.}\ }\textbf {\bibinfo {volume} {120}},\ \bibinfo {pages} {e2314212120} (\bibinfo {year} {2023}{\natexlab{a}})}\BibitemShut {NoStop}%
\bibitem [{\citenamefont {Wang}\ \emph {et~al.}(2023{\natexlab{b}})\citenamefont {Wang}, \citenamefont {Gupta}, \citenamefont {Singh}, \citenamefont {Madathil}, \citenamefont {Chung}, \citenamefont {Pfeiffer}, \citenamefont {Baldwin}, \citenamefont {Winkler},\ and\ \citenamefont {Shayegan}}]{Wang.PRL.2023}%
  \BibitemOpen
  \bibfield  {author} {\bibinfo {author} {\bibfnamefont {C.}~\bibnamefont {Wang}}, \bibinfo {author} {\bibfnamefont {A.}~\bibnamefont {Gupta}}, \bibinfo {author} {\bibfnamefont {S.~K.}\ \bibnamefont {Singh}}, \bibinfo {author} {\bibfnamefont {P.~T.}\ \bibnamefont {Madathil}}, \bibinfo {author} {\bibfnamefont {Y.~J.}\ \bibnamefont {Chung}}, \bibinfo {author} {\bibfnamefont {L.~N.}\ \bibnamefont {Pfeiffer}}, \bibinfo {author} {\bibfnamefont {K.~W.}\ \bibnamefont {Baldwin}}, \bibinfo {author} {\bibfnamefont {R.}~\bibnamefont {Winkler}},\ and\ \bibinfo {author} {\bibfnamefont {M.}~\bibnamefont {Shayegan}},\ }\bibfield  {title} {\bibinfo {title} {{Fractional Quantum Hall State at Filling Factor $\ensuremath{\nu}=1/4$ in Ultra-High-Quality GaAs Two-Dimensional Hole Systems}},\ }\href {https://doi.org/10.1103/PhysRevLett.131.266502} {\bibfield  {journal} {\bibinfo  {journal} {Phys. Rev. Lett.}\ }\textbf {\bibinfo {volume} {131}},\ \bibinfo {pages} {266502} (\bibinfo {year} {2023}{\natexlab{b}})}\BibitemShut
  {NoStop}%
\end{thebibliography}%

\clearpage
\onecolumngrid
{\section{End Matter}}

\twocolumngrid

In this End Matter (EM) we provide some notes clarifying the points made in the main text. 

\begin{enumerate}[label={[EM\arabic*]}]

    \item {The FQHSs at $\nu = 3/7$, 5/11, etc. are 1C because}, if they were 2C, then the layer fillings would be 3/14, 5/22, etc., which are not plausible FQHSs. Note also in Fig. \hyperlink{fig3}{2(a)} the emergence of a FQHS at $\nu=11/15$. This state has been reported previously in wide GaAs QWs at high densities, and was shown be to an “imbalanced” FQHS where the layer densities spontaneously become unequal \protect\cite{Manoharan.PRL.1997}.

    \item One would expect similar transitions for other even-numerator FQHSs near $\nu = 1/2$, but they are engulfed on the 2C side by the strong insulting phases that surround $\nu = 1/2$, especially for $\nu < 1/2$; see Ref. \cite{Manoharan.PRL.1996}.

    \item The transitions at $\nu = 2/3$ and 4/7 occur at slightly different values of $\alpha$. The 2/3 FQHS undergoes its 1C to 2C transition at $\alpha\simeq0.073$ as signaled by the narrowing of the 2/3 FQHS plateau.

    \item The blue, open square in Fig. \hyperlink{fig3}{2(b)} represents the density ($n\simeq 1.78$) at which, based on the phase diagram for the stability of 1/2 FQHS in wide GaAS QWs \cite{Shabani.PRB.2013}, we expect the 1/2 FQHS to collapse and give way to an insulating phase. In our experiments we do not have access to the magnetic fields required to reach $\nu= 1/2$ at $n=1.78$, but the weakening of the 1/2 FQHS and the dominance of the insulating phase as this density is approached is evident in Fig. \hyperlink{fig1}{1(d)}.

    \item The definitions of the tunneling terms in Refs. \cite{Peterson.PRB.2010, Zhu.PRB.2016} are different. In Ref. \cite{Peterson.PRB.2010} the authors define $t = \alpha$, whereas in Ref. \cite{Zhu.PRB.2016}, Zhu \textit{et al.} define the parameter $t_\perp = \alpha/2$. The data of Fig. 4(c) of Ref. \cite{Zhu.PRB.2016}, adapted as the red curve in Fig. \hyperlink{fig3}{2(b)}, has been appropriately scaled by the factor of 2 to match our definition of $\alpha$.

    \item The calculations in Ref. \cite{Peterson.PRB.2010} are done for a large range of parameters of layer thickness and $d/\ell_B$. Figure \hyperlink{fig3}{2(b)} is an example, with parameters closest to our sample’s parameters. See SM Section III \cite{SM.2025} for additional details.

    \item We add that in our experiments we do not see hysteresis or other signs of domain formation typically associated with a phase transition.

    \item {Given the indirect nature of our charge transport measurements,} we cannot rule out that the 1/2 FQHS remains 1C in the entire range of densities studied. This is an interesting possibility as it would imply that the 2DES is 2C over a large range of filling factors around $\nu = 1/2$, but the 1/2 FQHS remains 1C. {Detailed numerical calculations and direct measurements that probe the neutral transport properties can certainly shed more light.}

\end{enumerate}

\end{document}


\title{{Supplemental Material for ``Fractional quantum Hall state at $\nu = 1/2$ with energy gap up to 6 K, and possible transition from one- to two-component state"}}
\date{\today}
    
\author{Siddharth Kumar Singh, Chengyu Wang, Adbhut Gupta, Kirk W. Baldwin, Loren N. Pfeiffer, and Mansour Shayegan}
\affiliation{Department of Electrical and Computer Engineering, Princeton University, Princeton, New Jersey 08544, USA}

\maketitle

\tableofcontents

\clearpage

\setcounter{figure}{0}
\renewcommand{\figurename}{FIG.}
\renewcommand{\thefigure}{S\arabic{figure}}

\section{Determining $\Delta_{SAS}$}

{The 2D electrons in a wide GaAs qunatum well (QW) typically occupy the lowest two, symmetric and antisymmetric, electric subbands. The separation between these subbands, $\Delta_{SAS}$, is a crucial parameter in our study as it is a direct measure of the interlayer tunneling. We determine  $\Delta_{SAS}$ from the Fourier transform of the low-field Shubnikov-de Haas (SdH) oscillations \cite{Suen.PRL.1992, Suen2.PRL.1992, Suen.PRL.1994, Manoharan.PRL.1996, Shayegan.Review.LesHouches.1999}. The SdH oscillations of our 72.5-nm-wide sample at density, $n = 1.17$ (which we give here in units of $10^{11}$ cm$^{-2}$) are shown in Fig. \hyperlink{c3fig1}{S1(a)}. Note that the SdH oscillations are periodic in $1/B$. A close examination of the $R_{xx}$ data shows that the oscillations have a beating pattern which happens as $E_F$ moves between the Landau levels (LLs) corresponding to the symmetric and antisymmetric subbands. The Fourier transform of the $R_{xx}$ data of Fig. \hyperlink{c3fig1}{S1(a)} is shown in Fig. \hyperlink{c3fig1}{S1(b)}. Clear peaks in the Fourier transform spectrum are observed, and correspond to:

\begin{enumerate}
    \item $B_S$ - The frequency corresponding to the spin-unresolved, SdH oscillations of the electrons in the symmetric subband. 
    \item $B_{AS}$ - The frequency corresponding to the spin-unresolved, SdH oscillations of the electrons in the antisymmetric subband. 
    \item $B_S+B_{AS}$ - The frequency corresponding to the spin-unresolved, SdH oscillations of the electrons in the symmetric and antisymmetric subbands. This peak corresponds to the total density in the 2D electron system.
    \item $B_S-B_{AS}$ - The beat frequency corresponding to the spin-unresolved, SdH oscillations of the difference in the population of electrons in the symmetric and antisymmetric subbands. 
\end{enumerate}

The frequencies $(B_{S}\ {\rm and}\ B_{AS})$ obtained from the Fourier transform can be used to extract the corresponding densities $(n_{S} = eB_{S}/h)$ and $(n_{AS} = eB_{AS}/h)$, allowing us to extract $\Delta_{SAS} = (e/h)\times (B_{S} - B_{AS})/(m^*/\pi\hbar^2)$, where $m^*$ is the effective mass of electron in GaAs. Note that $m^*/\pi\hbar^2$ is the density-of-states for spin-unresolved 2D electrons. The extracted $\Delta_{SAS}$ as a function of \textit{n} are displayed in Fig. \hyperlink{c3fig1}{S1(c)}. The calculated charge distributions at two densities are shown as insets in Fig. \hyperlink{c3fig1}{S1(c)}.}

\begin{figure*}[t!]
\hypertarget{c3fig1}{}
\includegraphics[width=1\columnwidth]{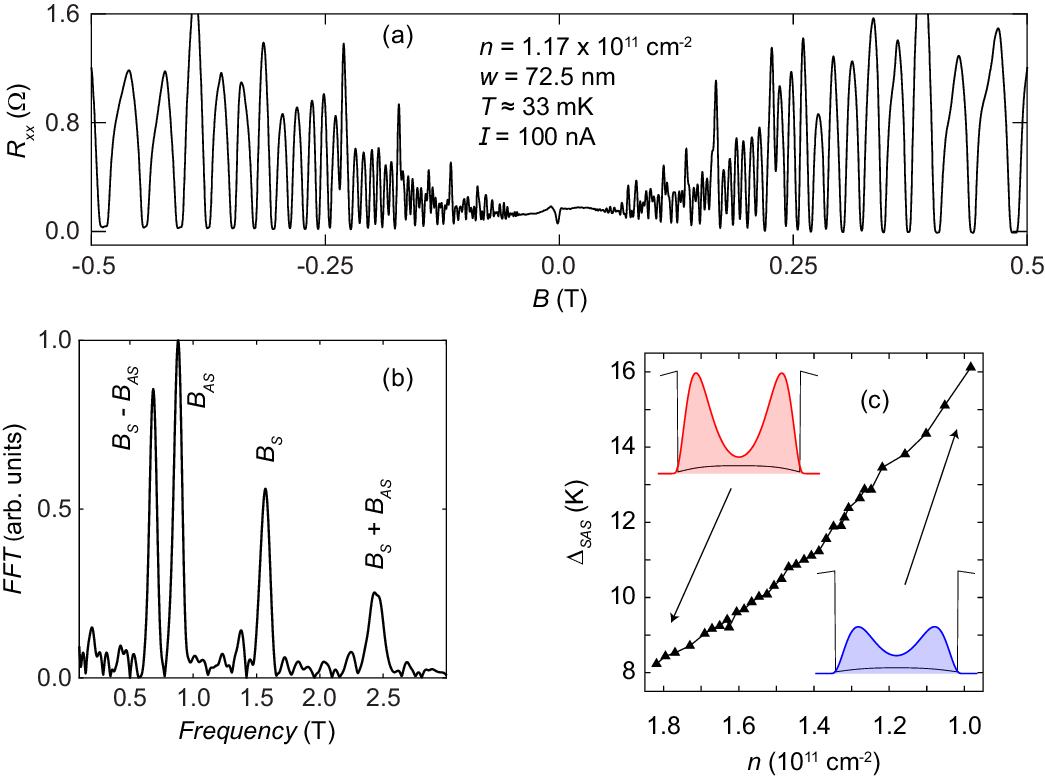}
\caption{(a) Shubnikov-de Haas (SdH) oscillations at $n = 1.17$. A large driving current $I = 100$ nA is used because of the small values of $R_{xx} \simeq 0.8-1.6\ \Omega$. The oscillations are periodic in $1/B$. (b) Fourier transform of the SdH oscillations shown in (a). Clear peaks in the Fourier transform can be seen at $B_{S}, B_{AS}, {\rm\ and}\ B_S \pm B_{AS}$; these are the frequencies of the spin-unresolved SdH oscillations. The Fourier transform peaks can be used to obtain the subband densities, the total density and $\Delta_{SAS}$. (c) $\Delta_{SAS}$ extracted from the SdH oscillations at low \textit{B}-field. Data are the same as Fig. 1(b) of the main text, reproduced here for simplicity. Self-consistent Hartree calculations of the charge distribution and potential at \textit{n} = 1.82 and 1.00 in a 72.5-nm-wide GaAs QW are shown as top and bottom insets, respectively.}
\end{figure*}

\section{Magneto-resistance traces for $0.98 \lesssim n \lesssim 1.82 \times 10^{11}$ cm$^{-2}$}

In Figs. \hyperlink{FigSM1-1}{S2-S3}, $R_{xx}$ traces of our 72.5-nm-wide GaAs sample are displayed for different \textit{n}. The traces are offset vertically for clarity. These $R_{xx}$ traces are used to construct the color-scale phase diagrams in Figs. 1(d) and 2(a). The rich physics of the 2DES in wide GaAs QWs is exemplified in these traces, with several correlated liquid and solid phases observed for $\nu<1$. 

\begin{figure*}[b!]
\hypertarget{FigSM1-1}{}
\includegraphics[width=1\columnwidth]{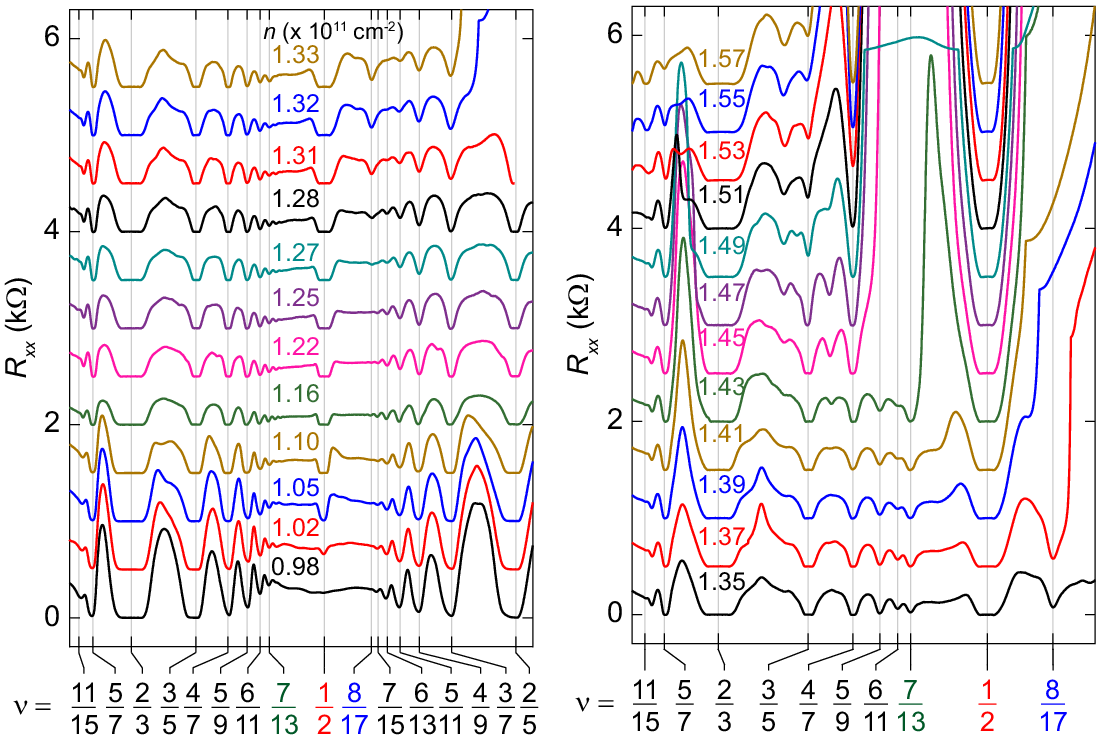}
\caption{$R_{xx}$ traces of our sample for $0.98 \lesssim n \lesssim 1.57 \times 10^{11}$ cm$^{-2}$. The daughter state of the 1/2 FQHS at $\nu = 8/17$ develops at lower densities compared to the one at $\nu = 7/13$ as is evident from the evolution displayed here. Also note the emergence of the bilayer Wigner crystal insulating phase for $n > 1.37$.}
\label{fig:figSM1}
\end{figure*}

\clearpage

\begin{figure*}[t!]
\hypertarget{FigSM1-2}{}
\includegraphics[width=1\columnwidth]{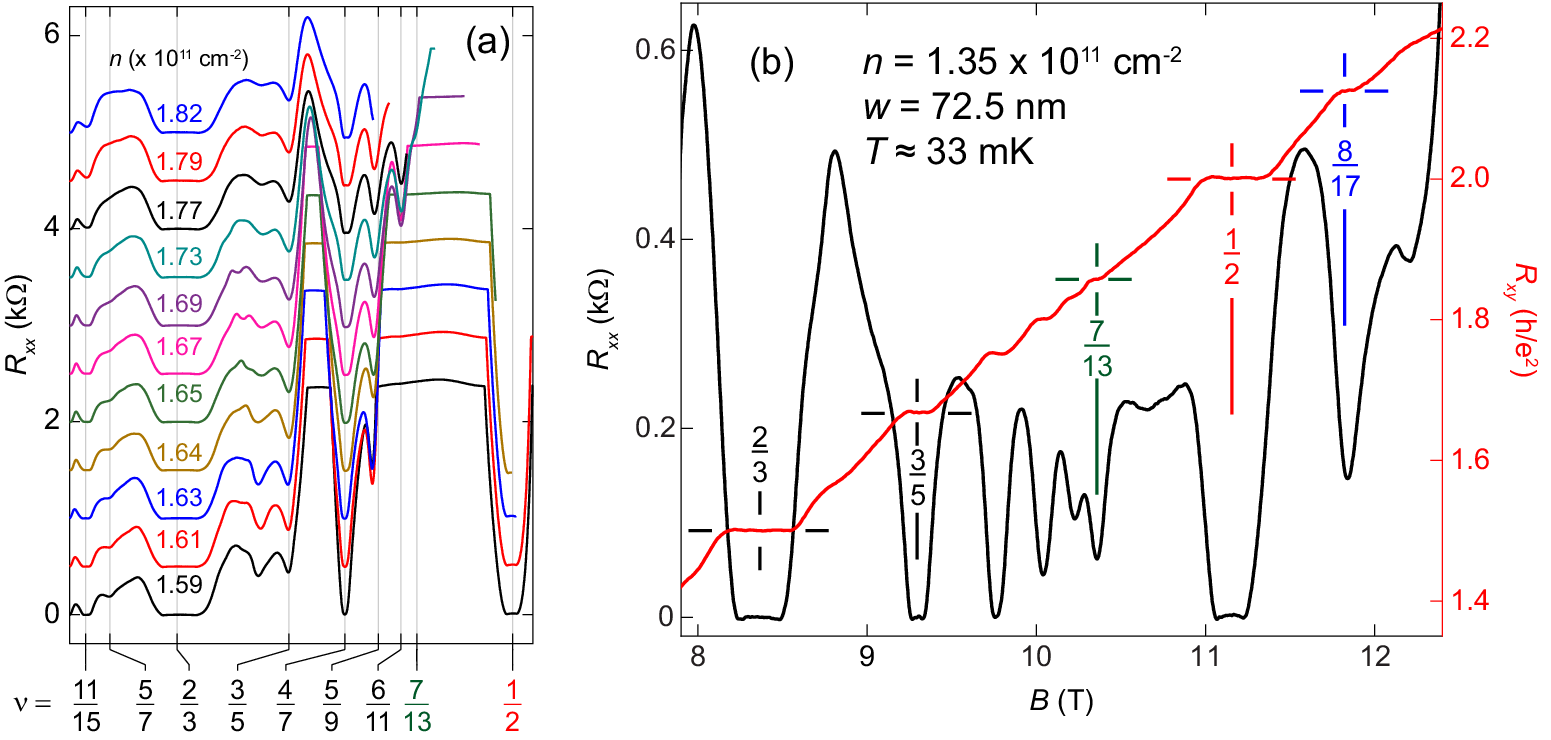}
\caption{(a) $R_{xx}$ traces of our sample for $1.59 \lesssim n \lesssim 1.82 \times 10^{11}$ cm$^{-2}$. (b) The daughter states of the 1/2 FQHS at $\nu = 8/17$ and 7/13 are fully developed at $n = 1.35 \times 10^{11}$ cm$^{-2}$. Well-developed plateaus in the $R_{xy}$ are observed for $\nu = 1/2$, 7/13 and 8/17.}
\label{fig:figSM1}
\end{figure*}

\section{Arrhenius plots for the extraction of the energy gaps of the 1/2 FQHS}

\begin{figure*}[t!]
\hypertarget{FigSM2}{}
\includegraphics[width=0.9\columnwidth]{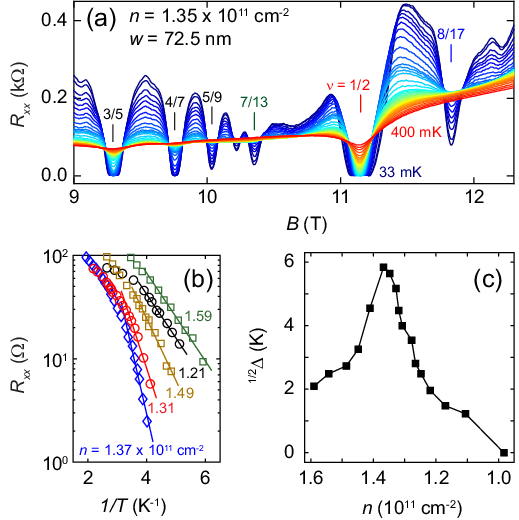}
\caption{(a) Temperature dependence of the $R_{xx}$ traces at $n = 1.35$ for $0.03 \lesssim T \lesssim 0.4$ K. (b) Arrhenius plots used to extract the energy gap of the 1/2 FQHS  ($^{1/2}\Delta$) for 5 different densities. (c) Extracted $^{1/2}\Delta$ [same as Fig. 1(e) of the main text, reproduced here for convenience].}
\label{fig:figSM2}
\end{figure*}

The energy gap $(^{1/2}\Delta)$ of the $\nu = 1/2$ FQHS quantifies the robustness of the state; larger $^{1/2}\Delta$ implies that the lowest lying charged excitations will be more resistant to external disturbances. To extract $^{1/2}\Delta$, several $R_{xx}$ traces are taken for $0.03 \lesssim T \lesssim 0.4$ K as shown in Fig. \hyperlink{FigSM2}{S4(a)}. The values of $R_{xx}$ minima at $\nu = 1/2$ are then fitted to the expression $R_{xx} \propto exp(-^{1/2}\Delta/2k_BT)$, where $k_B$ is the Boltzmann constant. To do this, the $R_{xx}$ minima are plotted as a function of $1/T$ in Arrhenius plots as shown in Fig. \hyperlink{FigSM2}{S4(b)}. Figure \hyperlink{FigSM2}{S4(b)} displays the Arrhenius plots used to extract $^{1/2}\Delta$ for 5 different densities. We emphasize that the range of temperature in which the 1/2 FQHS shows activated behavior is much larger than the melting temperature of the bilayer Wigner crystal. {This strongly suggests that the non-monotonic behavior of $^{1/2}\Delta$ as a function of \textit{n} (or $\alpha$) is an intrinsic property of the 1/2 FQHS itself, and not an artifact of the bilayer Wigner crystal.} The activation of even-denominator FQHSs can itself serve as an interesting topic as these are typically activated over a smaller range of temperatures compared to the Jain-sequence FQHSs \cite{Chung.NatMater.2021, Assouline.PRL.2024}. The measured $^{1/2}\Delta$ as a function of $n$ are shown in Figs. 1(e) and \hyperlink{FigSM2}{S4(c)}. 

\section{Sample parameters relevant to theoretical calculations}

\begin{figure*}[h!]
\hypertarget{FigSM3}{}
\includegraphics[width=1\columnwidth]{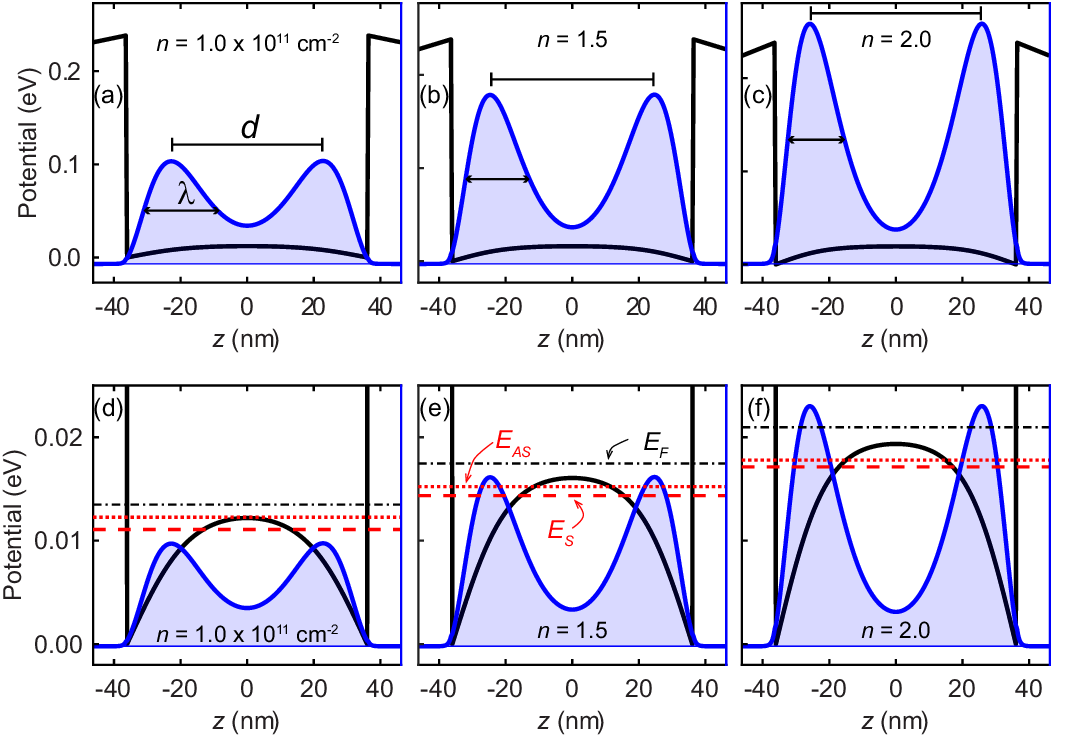}
\caption{(a)-(c) Results of self-consistent Schr\"odinger and Poisson (Hartree) calculations for the potential (black) and charge distribution (blue) in a 72.5-nm-wide GaAs QW for \textit{n} = 1.0, 1.5 and 2.0$\times10^{11}$ cm$^{-2}$, respectively. (d)-(f) The calculated potentials of (a)-(c) replotted on an expanded y-axis scale to show more clearly the changes in potential (solid black line), Fermi energy ($E_{F}$, dashed-dotted black line), and the energy levels of the symmetric ($E_S$, dashed red line) and antisymmetric ($E_{AS}$, dotted red line) subbands. The interlayer distance (\textit{d}) is depicted in (a)-(c) as the distance between the two peaks in the charge distribution. The ``layer thickness" $(\lambda)$ is defined as the full-width-at-half-maximum of each layer.}
\label{fig:figS2}
\end{figure*}

In this section, we describe in detail the parameters of the 2DES used in our experiments; see also Ref. \cite{Shabani.PRB.2013}. 

{Figure \hyperlink{FigSM3}{S5} displays the results of our self-consistent calculations of the potential and charge distribution in a 72.5-m-wide GaAs QW for three different electron densities. The height of the QW walls is set by the conduction-band offset of GaAs and Al$_x$Ga$_{1-x}$As, where $x$ is the alloy fraction of Al in the barrier ($x = 0.3$ for our sample). For the ungated case, the potential outside the QW is defined by the electric field generated by the positive charge of donor atoms (in our case, monolayer of Si) which are typically $>150$ nm far from the QW. {Our sample is fitted with back and front gates [see Fig. 1(a)], which allows us to change the electric fields on both sides of the QW and thus change the density.} The effect of the strength of the electric field is visible in the band bending, namely the slope of the potential, outside the QW walls; stronger electric fields result in larger 2DES density, as displayed in Figs. \hyperlink{FigSM3}{S5(a)-S5(c)}. 

The potential inside the QW is determined by the repulsion between the electrons. As the density is raised from $n = 1.0$ to $2.0$, the repulsion between the electrons raises the potential at the center of the QW, and pushes them towards the walls of the QW. Figures \hyperlink{FigSM3}{S5(d)-S5(f)} display the calculated potential on an expanded scale to show the behavior of the subband energies $(E_S, E_{AS})$ and the Fermi energy $(E_{F})$. The interlayer distance $(d)$ and the layer thickness $(\lambda)$ are indicated in Figs. \hyperlink{FigSM3}{S4(a)}-\hyperlink{FigSM3}{S4(c)}. The interlayer tunneling is determined by the energy difference of the lowest two electric subbands $(\Delta_{SAS} = E_{AS} - E_S)$. Note that as the density is raised, the subband energies begin to fall below the bump in the potential in the center of the QW.
    
\begin{figure*}[t!]
\hypertarget{FigSM4}{}
\includegraphics[width=1\columnwidth]{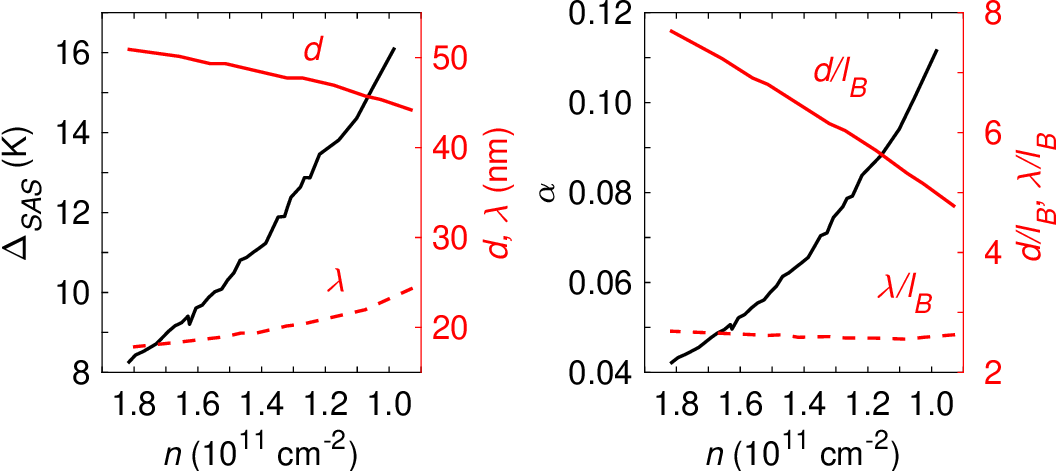}
\caption{The interlayer tunneling $(\Delta_{SAS})$, interlayer distance $(d)$ and the layer thickness $(\lambda)$ are plotted as a function of \textit{n} in absolute (left panel) and normalized units (right panel). The $\Delta_{SAS}$ data are deduced from the Fourier transforms of the low-field SdH oscillations in our sample as described in Section I, and the parameters $d$ and $\lambda$ are obtained through self-consistent Hartree calculations as shown in Fig. \protect\hyperlink{FigSM3}{S5}.}
\end{figure*}

The evolution of $d$, $\lambda$ and $\Delta_{SAS}$ with increasing \textit{n} is captured effectively in Fig. \hyperlink{FigSM4}{S6}. The parameters $\Delta_{SAS}$ (left panel) and $\alpha = \Delta_{SAS}/(e^2/4\pi\epsilon\ell_B)$ (right panel), plotted on the left axis, become smaller as \textit{n} is increased. Also shown are the parameters $d$ (solid, red curve) and $\lambda$ (dashed, red curve) on the right axis in Fig. \hyperlink{FigSM4}{S5} as a function of \textit{n}. Because of the very close competition between the one-component (Pfaffian), the two-component ($\Psi_{331}$) incompressible states and the compressible composite fermion (CF) Fermi sea (CFFS) state at $\nu = 1/2$ in wide GaAs QWs, the parameters $\Delta_{SAS},\ d$ and $\lambda$ play crucial roles in deciding which state has lower ground-state energy. These parameters determine the exact nature of the electronic interaction: whether the 2DES behaves as a single, thick, electron layer with softened short-range Coulomb interaction, or as a bilayer with significant layer thickness and interlayer tunneling.

\section{Comments on existing theoretical calculations}

Theoretical calculations of competing ground states in wide GaAs QWs have incorporated the role of finite layer thickness and interlayer distance to varying degrees \cite{Peterson.PRB.2010, Thiebaut.PRB.2015, Zhu.PRB.2016, Sharma.PRB.2024}. References \cite{Peterson.PRB.2010, Zhu.PRB.2016} only consider the one-component (1C) Pfaffian and the two-component (2C) $\Psi_{331}$ states, ignoring both the CFFS state and bilayer Wigner crystal states as competitors. As a result, their calculated phase diagrams of the stability of the 1/2 FQHS in wide GaAs QWs severely deviates from the experimentally-determined phase diagram \cite{Shabani.PRB.2013}. 

References \cite{Thiebaut.PRB.2015, Sharma.PRB.2024} consider the transition of the CFFS to a FQHS as a function of \textit{n} in wide GaAs QWs. They obtain a phase boundary between the CFFS and the FQHS at $\nu = 1/2$ in the well-width (\textit{w}) vs 2DES density (\textit{n}) phase space; both calculated phase boundaries are in reasonably good agreement with the experimental phase boundary reported in Ref. \cite{Shabani.PRB.2013}. Reference \cite{Thiebaut.PRB.2015} additionally calculates the phase boundary between the 1/2 FQHS and the insulating phase, which also agrees reasonably well with the experimental phase boundary. 

Reference \cite{Thiebaut.PRB.2015} finds that as the density is raised, the CFFS transitions into the 2C, Abelian, $\Psi_{331}$ state. However, Ref. \cite{Thiebaut.PRB.2015} approximates the charge distribution of the 2DES by assuming that the lowest two electric subbands have a \textit{sine} and \textit{cosine} functional form. As noted above, the exact nature of the electronic interaction is of paramount importance in deciding the ground-state energy of the different competing states. 

In contrast, Sharma \textit{et al.} \cite{Sharma.PRB.2024} demonstrate that the CFFS transitions into a 1C incompressible state described as a \textit{p-wave} paired state of CFs. This state is in the same universality class as the 1C, non-Abelian, Pfaffian state. Sharma \textit{et al.} calculate the charge distribution of the 2DES confined to a wide GaAs QW using local density approximation (LDA). The LDA calculations should accurately capture the charge distribution in our samples, especially in the low density regime where $\Delta_{SAS}$ is large. Our experimental findings suggest that the CFFS at $\nu = 1/2$ transitions to a 1C FQHS, consistent with the \textit{p-wave} paired state of CFs discussed in Ref. \cite{Sharma.PRB.2024}.}

{
\section{Theoretical phase boundary for the 1C to 2C transition of the 1/2 FQHS}

\begin{figure*}[b!]
\hypertarget{FigSM_gaps}{}
\includegraphics[]{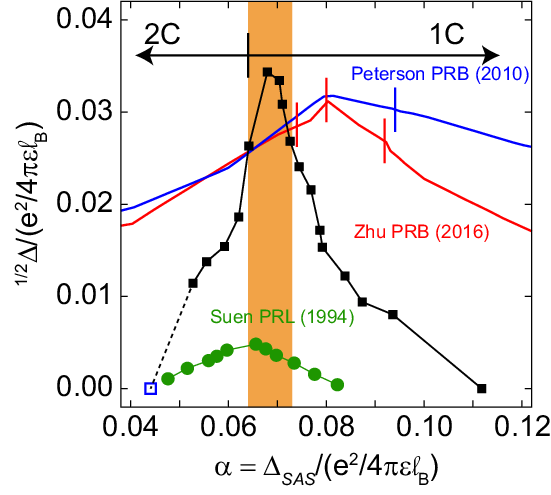}
\caption{Figure 2(b) of the main text is reproduced here, for simplicity. Solid black squares are the experimentally measured energy gap of the 1/2 FQHS. The open, blue square is obtained from the phase diagram of the stability of the 1/2 FQHS in Ref. \cite{Shabani.PRB.2013}, and denotes the point at which the bilayer Wigner crystal insulating phase takes over at $\nu = 1/2$. The green, circles are measurements of the energy gap of the 1/2 FQHS by Suen \textit{et al.} \cite{Suen.PRL.1994}, in samples with 8 times smaller mobility. The blue and red curves are theoretical calculations taken from Ref. \cite{Peterson.PRB.2010} and Ref. \cite{Zhu.PRB.2016}, respectively. The positions of the 1C to 2C transition are marked with short, vertical lines.}
\end{figure*}

Figure \hyperlink{FigSM_gaps}{S7} displays the experimentally measured energy gaps of the 1/2 FQHS (solid black squares) plotted as a function of the parameter $\alpha$. The open blue square represents the point at which the 1/2 FQHS is expected to be destabilized and the bilayer Wigner crystal should take over. Because we are limited by the magnetic field available to us in our experiments, we have adopted the open blue square from the phase diagram of Ref. \cite{Shabani.PRB.2013}. The orange band represents the region of stability of the 8/17 and 7/13 daughter FQHSs of the 1/2 FQHS. The left edge of the orange band is at $\alpha \simeq 0.064$ and marks the position of the 1C to 2C transition of the 2DES as determined from: (i) The appearance of the bilayer Wigner crystal insulating phases flanking the 1/2 FQHS. The bilayer Wigner crystal also destabilizes the Jain FQHSs as well as the daughter FQHSs of the 1/2 FQHS. (ii) The disappearance of the odd-numerator FQHSs at $\nu = 3/5$ and 5/7. (iii) The appearance of the imbalanced, bilayer FQHS at $\nu = 11/15$. Other hints of the 1C to 2C transition of the 1/2 FQHS can be seen at even-numerator FQHSs at $\nu = 2/3$ and 4/7, but these occur at slightly different values of $\alpha$. The black vertical line on the left edge of the orange band is used to  denote the 1C to 2C transition of the 2DES. It should be kept in mind that the 1C to 2C transition of the 1/2 FQHS may not coincide with that of the 2DES, or at other $\nu$. However, the main purpose of our study is to provide evidence for the existence of such a transition. We leave the exact position of this transition to future studies. 

The blue curve in Fig. \hyperlink{FigSM_gaps}{S7} depicts the theoretically calculated energy gap of the 1/2 FQHS as reported in Fig. 6(b) of Peterson and Das Sarma \cite{Peterson.PRB.2010}. In Ref. \cite{Peterson.PRB.2010}, the authors define the parameter $t$, which they use in their calculations to account for the interlayer tunneling. In the definition used by Peterson and Das Sarma, the parameter $t = \alpha$. Peterson and Das Sarma perform numerical exact diagonalization of $N = 8$ electrons to obtain the ground state at $\nu = 1/2$. The overlaps of the numerically-calculated ground state at $\nu = 1/2$ with the Pfaffian and the $\Psi_{331}$ states are then used to determine the value of $\alpha$ at which the transition of the 1/2 FQHS occurs from the Pfaffian to the $\Psi_{331}$ state. In Fig. \hyperlink{FigSM_gaps}{S7} we mark their reported value forthe transition by a blue vertical line.

The red curve in Fig. \hyperlink{FigSM_gaps}{S7} features the theoretically calculated energy gap of the 1/2 FQHS as reoported in Fig. 4(c) of Ref. \cite{Zhu.PRB.2016}. Note that in Fig. 4(c) of Ref. \cite{Zhu.PRB.2016}, the authors plot the calculated energy gap of the 1/2 FQHS as a function of the parameter $t_\perp$. In the definition used by Zhu \textit{et al.}, $t_\perp = \alpha/2$, and we have appropriately accounted for this factor of 2 in our reproduction shown in Fig. \hyperlink{FigSM_gaps}{S7}. Zhu \textit{et al.} identify signatures of the transition of the 1/2 FQHS from the non-Abelian, Pfaffian state to the Abelian, $\Psi_{331}$ state using three different theoretical calculations, and report three different values of $\alpha$ for the transition of the 1/2 FQHS (these are marked by short red vertical lines): (i) The ground state energy spectra reveal a change of the ground state degeneracy at $\alpha \simeq 0.092$, (ii) The maximum in the charge transport energy gap of the 1/2 FQHS at $\alpha \simeq 0.08$, and (iii) Discontinuity in the second derivative of the ground state energy of the 1/2 FQHS at $\alpha \simeq 0.074$. 

}

\section{Comments on measured 1/2 FQHS energy gaps}

Finally, the $^{1/2}\Delta$ reported by Suen \textit{et al.} \cite{Suen.PRL.1994} also show a similar trend - $^{1/2}\Delta$ increases as $\alpha$ is reduced, reaches a maximum and then begins to decrease. However, the gaps measured in Ref. \cite{Suen.PRL.1994} are smaller than ours by almost an order of magnitude, and are therefore strongly affected by disorder. The weaker 1/2 FQHS also becomes much more susceptible to the effect of disorder and the competing bilayer Wigner crystal on its flanks. In our data, we observe a strong 1/2 FQHS in the entire range of $n \gtrsim 1.05$ (see Figs. \hyperlink{FigSM1-1}{S2-S3}). The disorder parameter $(\Gamma)$ in our sample is $\simeq$ 0.5 K \cite{VillegasRosales.PRL.2021, Singh.NatPhys.2024} which is much smaller than $^{1/2}\Delta$ in our sample in a large density range [see Figs. \hyperlink{SM2}{S4}, 1(e)]. The much lower level of disorder in our sample is also very likely responsible for the enhanced robustness of the 1/2 FQHS with charge distribution asymmetry compared to Ref. \cite{Suen.PRL.1994} sample. 

\bibliography{aps}